\documentclass[preprint,amsmath,noshowkeys,noshowpacs,longbibliography,amssymb,aps,pra, floatfix]{revtex4-1}

\usepackage{graphicx,color}
\usepackage{cleveref}
\usepackage{amsthm,mathrsfs,amsopn}
\usepackage{mathtools}

\usepackage{cases}
\usepackage{booktabs}
\usepackage{bbm}
\usepackage{bm}

\usepackage[caption=false]{subfig}

\begin{document}
\raggedbottom

\title{Train Stochastic Non Linear Coupled ODEs to Classify and Generate}

\author{Stefano Gagliani $^1$, Feliciano Giuseppe Pacifico $^{1,2}$, Lorenzo Chicchi $^1$, Duccio Fanelli $^1$, Diego Febbe $^1$, Lorenzo Buffoni $^1$, Raffaele Marino $^1$}

\affiliation{$1$ Department of Physics and Astronomy, University of Florence, Sesto Fiorentino, Italy \\ INFN, Italy}
\affiliation{$2$ Department of Informatics and Computer Science, University of Pisa, Italy }

\begin{abstract}
A general class of dynamical systems which can be trained to operate in classification and generation modes are introduced. A procedure is proposed to plant asymptotic stationary attractors of the deterministic model. Optimizing the dynamical system amounts to shaping the architecture of inter-nodes connection to steer the evolution towards the assigned equilibrium, as a function of the class to which the item – supplied as an initial condition - belongs to. Under the stochastic perspective, point attractors are turned into probability distributions, made analytically accessible via the linear noise approximation. The addition of noise proves beneficial to oppose adversarial attacks, a property that gets engraved into the  trained adjacency matrix  and therefore also inherited by the deterministic counterpart of the optimized stochastic model. By providing samples from the target distribution as an input to a feedforward neural network  (or even to a dynamical model of the same typology of that adopted for classification purposes), yields a fully generative scheme. Conditional generation is also possible by merging classification and generation modalities. Automatic disentanglement of isolated key features is finally proven.  
\end{abstract}

\maketitle
\newpage
\section{Introduction}

Deep Neural Networks (DNNs) represent the current benchmark for regression and classification across a wide range of disciplines \cite{bishop2023deep}. Their ability to identify complex patterns and correlations places them at the forefront of advanced artificial intelligence systems. Scalar weights, which link adjacent nodes across feed-forward architectures, follow an optimization algorithm and store the information needed for the trained network to perform the assigned tasks with unprecedented accuracy and fidelity. The modern evolution of deep neural networks, the so-called transformer-based models, proves extremely versatile in dealing with intricate data sequences for  multimodal usage, attaining cutting-edge advancements across nearly every machine learning discipline, such as computer vision, speech recognition and time series forecasting \cite{vaswani2017attention, gunasekaran2025predictive, kalluri2025computer, liu2025machine}. In the context of transformers, feed-forward modules hide as nested sub-units within each encoder and decoder layer, and transform further the output from the self-attention layer. During training stages, the model iteratively quantifies the prediction errors with an appropriate loss function, and updates its parameters (weights and biases) through backpropagation and gradient descent to learn patterns from the data. Also in this framework, of current and widespread modern usage, the goal is again to adjust billions of parameters, including weights in attention blocks and feed-forward layers, until the model reliably produces accurate outputs. The {\it network intelligence} - as one can figuratively refer to the acquired ability to learn, make predictions, and perform complex tasks - is ultimately encoded in the immense collection of internal parameters, weights and biases, which get adjusted during training to minimize errors, enabling the network to recognize patterns, make decisions, and generalize from data it has never seen before. For this reason, and despite their proven practical success, several theoretical aspects of neural networks remain poorly understood. Dissecting trained model to look for prototypical signatures of seemingly intelligent behaviors is extremely challenging, if not prohibitive, due to opaque imprint left by the empirically determined parameters on the ensuing patterns of activation\cite{rudin2019stop}. Feedforward modules, which fuel any deep learning solutions, from basic to complex architectures, can indeed approximate any continuous function on a compact subset of the real numbers. Their impersonal formulation in terms of alternate linear and non-linear transformations and the circumstantial lack of a physical interpretation for the involved parameters are among the factors that limit the contextual exploration via purely mathematical means. 

On a different level of abstraction, the brain, whose functions one seeks to eventually emulate through plausible digital analogues, can be pictured as a highly advanced dynamical system. It receives information from the senses and generates complex outputs that reflect the continuously changing, time-dependent patterns of the emitted neural activity. A rich plethora of computational neuroscience models exist that aim at reproducing, with various degree of sophistication, the dynamics underlying the actual brain functioning \cite{coombes2023neurodynamics}. The studied model consists of closed sets of (ordinary or stochastic) differential equations, encompassing different spatial and temporal scales, and triggering distinct macroscopic self-organized responses, as following the processed stimuli. Local and non-local interactions among neurons, the backbone of sensible models of the brain, are not delegated to implicit approximants but explicitly accommodated for, following an informed mathematical translation of biologically sensible mechanisms. Parameters involved are meaningful in terms of their associated biological connotation, contrary to what happens in over-parameterized neural networks. As stated above, in this latter setting and for any possible level of architectural complexity adopted, parameters are just input of feed-forward modules and it is hard to infer a posteriori their actual relevance within the redundant constellation that they happen to populate. Starting from these premises, training dynamical models suited for computational neuroscience could contribute to uncover new theoretical frameworks and elaborate on the fundamental mechanisms underlying deep neural network operational under a bio-mimetic perspective. 

Among existing models, recurrent neural networks (RNNs) are particularly notable, having demonstrated strong performance in time series analysis and natural language processing. RNNs are characterized by an internal state vector that is updated iteratively alongside the input, producing an output and simultaneously evolving the internal state. In this formulation, network depth corresponds to the number of iterations, effectively repeating the same computation through shared parameters. Time, flowing across layers, regains thus the stage it deserves as a fundamental ingredient under a compelling dynamical angle. Nonetheless, RNNs cannot be written in terms of  
closed differential equations of the type employed in any conceivable model. As such, they materialize as perfectly performing black boxes, but impervious to mathematical core drilling \cite{mienye2024recurrent}.

A significant generalization of this concept was introduced in \cite{chen2018node} with the concept of Neural Ordinary Differential Equations (nODEs), where the evolution of the internal state is governed by a parameterized differential equation. This can be interpreted as the continuous-time limit of a RNN: nODEs exemplify how dynamical systems can be employed to tackle standard learning tasks, achieving results comparable to conventional DNNs. Still, nODEs make use of implicit feedforward modules and this cast obvious limitation to our ability to probe their functioning with usual techniques borrowed from the theory of dynamical systems.

A variant of this framework, tailored for classification tasks and named SA-nODEs has been recently introduced in \cite{marino2023sanode}. This model consists of $N$ linearly coupled nodes, each evolving within a double-well potential landscape. Unlike standard nODEs,  the model is equipped with a set of stable attractors that become the target of the learnable dynamics, while the Euler numerical integration outlines the RNN architecture. This allows in turn to perform a straightforward optimization of the residual parameters for the system to react differently from distinct classes of processed data. More specifically, input items belonging to a given category, and supplied to the dynamical model as an initial condition, are directed towards a pre-designed equilibrium, that can be made stable either analytically via \textit{a priori} constraints or by means of the optimization procedure. Remarkably, for what it is here of interest, the trained model can be explicitly written in terms of an ensemble made of coupled ordinary differential equations, with no implicit approximants - namely, feed-forward units - involved. This paves the way to opening the black box of deep neural networks by 
conventional analytical means so as to cast novel light onto the implemented decision making process, in relation to the examined task. Inspired by biological models, preliminary  analysis along the same pathway, have been also conducted for the celebrated Continuous-Variable Firing Rate model \cite{kim2019, rungratsameetaweemana2025random}, a paradigmatic model of interacting non linear neurons widely employed in computational neuroscience applications.
  
Beginning with these ideas, we take here a decisive leap forward and provide a universal recipe to tailor suited dynamical systems - including those relevant to neuroscience applications - for classification and generation purposes. In particular, we introduce a numerically robust recipe to plant (deterministic or stochastic) attractors. The properties of the stochastic system prove rather insightful to developing a dynamically assisted generative scheme, starting from a basic reconstruction setting. The supplied initial condition, e.g. an image, is stored in the system correlations. In other words, it is not only possible to train a simple neural network that reconstructs the original image from its late-time latent representation, but also to generate new images by feeding values sampled from the (almost) asymptotic distribution. Automatic disentanglement of isolated key features is yet another consequence inherited by the dynamical nature of the algorithm operated under generative modalities. In the following section, we shall begin by introducing the reference mathematical framework and outline the universal character of the proposed scheme. 

\section{The mathematical framework}

Consider a population of $N$ interacting nodes and assume each node to be assigned a scalar dynamical variable $x_i$, with $i$ running from $1$ to $N$. Then, we posit:

\begin{equation}
\label{general_model}
\dot{x}_i = f(x_i) + \sum_j A_{ij} g(x_i, x_j) + \epsilon \sum_j G_{ij} \eta_j,  
\end{equation}

where a dot placed above the function name denotes the time derivative of that function; $f(\cdot)$ and $g(\cdot)$ are generic non linear functions of the state of the system; $\eta_i$ are delta correlated normally distributed random variables; $A_{ij}$ are the entries of the adjacency matrix $A$, which sets the web of inter-tangled binary interactions; matrix $G$, of elements $G_{ij}$, defines the correlation of the stochastic drive, while $\epsilon$ sets the strength of the imposed noise. In the following, we will prove that (i) it is always possible to a priori enforce an arbitrary number of suited stationary solutions of model (\ref{general_model}), in the deterministic limit $\epsilon=0$; (ii) for sufficiently small $\epsilon$, the system admits an equal set of stochastic attractors which can be analytically characterized as multivariate Gaussian, under the linear noise approximation; (iii) the elements of $A$ (and $G$, depending on the adopted settings) can be trained at will to direct different input items towards distinct asymptotic attractors, conditional to the specific class of pertinence; (iv) consider further the trained dynamical model operated as a sort of auto-encoder (thus, when forced to reproduce as an output the supplied input, via a complementary neural network):  items that happen to share common traits, while still belonging to just one isolated class are spontaneously fragmented in separated, though contiguous, sub-groups, in latent space. These latter sub-groups appear at late time (right before the asymptotic convergence, when spatial correlations fades eventually away) as distinct self-organized 'bubbles' (or regions) within the embedding space, smoothed by noise. The inherent ability displayed by the trained dynamical system to produce a hierarchically structured organization of the processed data is crucial for an effective generation scheme. Points (i) and (ii) as listed above will be discussed with reference to the general framework (\ref{general_model}). To elaborate on points (iii) and (iv) we will instead assume $f(x_i)=-r_i x_i$ and $g(x_i, x_j) = \beta x_i^2 / (c + x_i^2)$ as implemented in \cite{chicchi2025deterministic}. This implies in turn working with a particular version of the Continuous-Variable Firing Rate. The conclusions reached apply however in general for any possible choices of the model functions that falls under the descriptive umbrella of equations (\ref{general_model}).

\section{Planting stationary attractors: the case $\epsilon=0$.}
\label{subsec:Planting the deterministic attractors}

We shall here discuss the limiting setting $\epsilon=0$ and depict a general procedure to plant a set $C$ stationary solutions, denoted $\bar{x}^{(\ell)}$ with $\ell=1,...,C$, within a general model of the type postulated in equation (\ref{general_model}). To this end we posit:

\begin{equation}
\label{defA}
    A = \tilde{A} (1-P)+\tilde{P},
\end{equation}
where $\tilde{A}$ is a $N \times N$ matrix that is linked to $A$ by the above transformation. Matrix $\tilde{A}$ defines the target of the training as we will outline below. In the above equation (\ref{defA}):
 
\begin{align}
    P &= \sum_{l}\frac{g(\bar{x}^{(\ell)})g(\bar{x}^{(\ell)})^T}{||g(\bar{x}^{(\ell)})||^2},\label{ProjOp}\\
    \tilde{P}&=- \sum_{l}\frac{f(\bar{x}^{(\ell)})g(\bar{x}^{(\ell)})^T}{||g(\bar{x}^{(\ell)})||^2}.
\end{align}

Further, we require that the $C$ target points $\bar{x}^{(\ell)} \in \mathbb{R}^N$ are chosen such that the $g(\bar{x}^{(\ell)})$ are all mutually orthogonal. Under this condition, it is straightforward to show that $\bar{x}^{(\ell)}$ is a stationary solution of equations (\ref{general_model}) for the limiting setting where the dynamics is turned deterministic, i.e. by positing $\epsilon=0$. 

As an additional accessory requirement, one can demand that the images through $g(\cdot)$ of the aforementioned attractors are eigenvectors of the adjacency matrix $A$:
\begin{equation}
    A g(\bar{x}^{(\ell)})=\lambda_\ell g(\bar{x}^{(\ell)}),
\end{equation}

Recalling condition (\ref{defA}), this readily implies:

\begin{equation}
\label{cond_alphabet}
f(\bar{x}_i^{(\ell)})+\lambda_\ell g(\bar{x}_i^{(\ell)})=0,
\end{equation}

a constraint that should be matched by the individual components ${x}_i$, $i=1,...,N$ for $g(\bar{x}^{(\ell)})$ to materialize as an eigenstate of the coupling matrix $A$. The potential interest of this additional constraint resides in that condition (\ref{cond_alphabet}) yields a finite set of eligible values (the actual solution of the above equation) from which the equilibria $\bar{x}^{(\ell)}$ can be built. Also, and as we shall discuss in the following, linear stability analysis around the planted attractors can be explicitly carried out, under these premises.

If we further assume $g(0)=f(0)=0$, then $\bar{{x}_i}^{(\ell)}=0$ belongs to the alphabet of digits which can be used for constructing the sought solutions $\bar{x}^{(\ell)}$ and this makes it straightforward to impose the  orthogonality of $g(\bar{x}^{(\ell)})$ (note that assumption $g(0)$ suffices for handling the orthogonality condition without resorting to the above explicit alphabet). Denote by $a$ a non trivial solution of the non linear equation (\ref{cond_alphabet}). Then, for the orthogonality of $g(\bar{x}^{(\ell)})$ to be matched, it is sufficient to alternate $0$ and $a$ in the definition of each individual $\bar{x}^{(\ell)}$ in such a way that $x^{(\ell^*)}_i=a$ for a given $\ell^*$, and  $\bar{x}^{(\ell)}_i=0$, for $\ell \ne \ell^*$. By resorting to a minimal alphabet of two letters, $0$ and $a$, one can therefore craft at will the attractors of the dynamical model (\ref{general_model}), subject to the aforementioned orthogonality constraint. The employed alphabet can be made larger as reflecting the existing set of distinct solutions of the non linear equation (\ref{cond_alphabet}). 

As a notable case to be thoroughly analyzed, we set $\bar{x}^{(\ell)}_i= a \delta_{i\ell}$. This amounts to assume that the stationary states $\bar{x}^{(\ell)}$ align to the first $C$ elements of the canonical basis. Under this assumption, and having set $g(\bar{x}^{(\ell)})$ to unfold as eigenvectors of matrix $A$, one can carry out explicitly a linear stability analysis of the imposed equilibria. To this end we put forward the additional assumption $g'(0)=0$, where the prime symbol stands for the derivative operation. Impose a slight perturbation $\delta x_i$ around the  stationary solution $\bar{x}^{(\ell)}$ , namely $x_i = \bar{x}^{(\ell)}_i + \delta x_i$. Plugging into the governing equations  
(\ref{general_model}) and expanding at the linear order yields:

\begin{equation}
\label{ls1}
    \delta \dot{x}_i = f'(\bar{x}^{(\ell)}_i) \delta x_i + g'(a)\sum_{j \, \text{s.t.} \, \bar{x}^{(\ell)}_j \ne 0}^N A_{ij}\delta x_j,
\end{equation}
where $\bar{x}^{(\ell)}_i$ takes values $0$ or $a$. Recall that the non linear image of the first $C$ canonical elements are eigenvectors of $A$, and thus $A_{i\ell} = \lambda_\ell \delta_{i\ell}$. Hence,
equation (\ref{ls1}) can be cast in the form:

\begin{equation}
\label{ls2}
    \delta \dot{x}_i = \left( f'(\bar{x}^{(\ell)}_i) + g'(a) \lambda_\ell\right) \delta x_i. 
\end{equation}

Denote $\gamma = \text{max} (0,f'(a))$, than 
$\gamma + g'(a) \lambda_\ell <0$ is the condition that should be matched by the free parameters $\lambda_\ell$, the eigenvalues of $A$ relative to the eigenvectors $g(\bar{x}_i^{(\ell)}) \equiv \delta_{i\ell} g(a)$, for linear stability of $\bar{x}^{(\ell)}$ to hold.

To make more accessible the analysis carried out above, we make reference to the Continuous-Variable Firing Rate model. This latter scheme falls within the realm of (\ref{general_model}) with a specific choice of the non linear functions $f(\cdot)$ and $g(\cdot)$, as recalled in the end of the preceding Section (here $r_i$ is set to one). We notice in particular that $f(0)=g(0)=0$. The non trivial solutions of equation (\ref{cond_alphabet}) read:

\begin{equation}
    \bar{x}^{(\ell)}_i\equiv a_\pm = \frac{\beta\lambda_\ell \pm\sqrt{\beta^2\lambda_\ell^2-4c}}{2}\label{alphabet}.
\end{equation}

Assume now that $a=a_+$. We further notice that $g'(0)=0$, as requested for the linear stability analysis to hold. Also $\gamma=-1$ and  $g'(a)=2 c  g^2(a)/a^3$. Thus, in this specific case, the general condition for linear stability returns 
$c<\lambda_\ell \beta a/2$, where in deriving the above result use has been made of the self-consistent relation $\lambda_\ell g(a)=\bar{x}$ that follows constraint (\ref{cond_alphabet}) 
for $\bar{x}_i^{(\ell)}=a$. Remarkably the above condition is straightforwardly met, if $c<\frac{\beta^2 \lambda^2_\ell}{4}$ meaning that $a_\pm$ are real.    

At this point, it should be noted that the recipe to impose stationary solutions via the projection operator shares some similarity with the usual implementation of Hebbian learning \cite{amit1989, hebb2005} in Hopfield Networks \cite{hopfield1984graded}. While our task has little to do with modeling associative memory, we still need to address the presence of spurious attractors once the $\bar{x}^{(\ell)}$ are being selected. In the literature this is fundamental for quantifying the learning capacity of a network. At variance,  we are not here interested in maximizing $C$, but rather in minimizing spurious stable basins of attraction that could hinder the designated target attractors (or, alternatively, in exploiting the spurious attractors as designated targets of the dynamics). The number of existing attractors can be exponential in the number of stored ones, and the particular choice of the non linear functions $f(\cdot)$ and $g(\cdot)$ plays a role in this respect \footnote{As a basic example, imagine $f(\cdot)$ to be linear and choose an odd non linear function $g(\cdot)$. Then for each $\bar{x}^{(\ell)}$, the negation $-\bar{x}^{(\ell)}$ comes as a spurious attractor.}.

To elaborate further on these aspects, assume:

\begin{equation}
    \bar{x}^{(\ell)}_i=\begin{cases}
        a_+\ \ \ \text{if }\ \   i=(l-1)L+1, \dots, lL\\
        0 \ \ \ \ \ \text{otherwise} ,
    \end{cases}\label{AttrChoice}
\end{equation}
where $L = \lfloor \frac{N}{C}\rfloor$, with $N$ dimension of the latent space. Such a configuration, along with all the equivalent combinations of the $\ell$ non-zero components for each $\bar{x}^{(\ell)}$, has the immediate downside of generating an exponential amount of attractors. In fact, one can see that we inadvertently planted the following set of attractors:
\begin{equation}
    \sum_\ell c_\ell \bar{x}^{(\ell)} \ \ \ \ \ \text{with } c\in\Big\{0, 1, \frac{a_-}{a_+}\Big\}, \label{spurious}
\end{equation}
resulting in $3^C$ stationary solutions \footnote{It is important to stress that these figures are just lower bounds estimates. The phase space varies throughout the training stages, making it hard to come out with definite a priori assessments on the number of possible stationary states.}. The above strategy (\ref{AttrChoice}) will be implemented throughout the analysis reported above. In Appendix \ref{App1} we will elaborate further on the attractors' degeneracy, with reference to the classification task.

\section{The stochastic version of the model ($\epsilon \ne 0$): working under the linear noise approximation}

We shall here turn to discussing the general setting $\epsilon \ne 0$, under the accessory assumption $\epsilon << 1$. Focus on the dynamics around the generic attractor $\bar{x}^{(\ell)}$ and posit:

\begin{equation}
\label{VKansatz}
{x}_i = \bar{x}^{(\ell)}_i + \epsilon \xi_i ,
\end{equation}

where $\xi_i$ stands for the stochastic correction, order $\epsilon$, around the deterministic solution. Plugging (\ref{VKansatz}) into the ruling equations (\ref{general_model}), expanding in $\epsilon$ and keeping the leading terms yields the following linear Langevin equations:

\begin{equation}
\dot{\xi_i} = \sum_j J_{ij} \xi_j + \sum_j G_{ij} \eta_j,
\end{equation}

where  $J_{ij}= \left( f'(\bar{x}^{(\ell)}_i) \delta_{ij} + g'(\bar{x}^{(\ell)}_j) \right)$ are the elements of the Jacobian matrix $J$ (note that $J$ depends on the specific attractor $\ell$, even though this is not formally stressed for the ease of notation). The above Langevin equation yields the following Fokker-Planck equation 

\begin{equation}
\frac{\partial}{\partial t}  \Pi(\xi,t) = - \sum_i \frac{\partial}{\partial \xi_i}  \left( A_{i} \Pi(\xi,t) \right ) + \frac{1}{2} \sum_{i,j}
\frac{\partial^2}{\partial \xi_i \partial \xi_j} \left( D_{ij} \Pi(\xi,t) \right)
\end{equation}

for the evolution of the distribution of fluctuations $\Pi(\xi,t)$, with $\xi=\left(\xi_1, \xi_2, ..., \xi_i, ..., \xi_N \right)$, around the selected deterministic attractor $\bar{x}^{(\ell)}$. In the above equation, $A_{i} = \sum_j J_{ij} \xi_j$ and $D = GG^T$. The solution of the Fokker-Planck is a multivariate Gaussian:

\begin{equation}
\Pi(\xi,t) = \frac{1}{\sqrt{\left(2 \pi \right)^N \det(\Sigma) }} 
\exp \left( -\frac{1}{2} \xi^T \Sigma_\ell^{-1} \xi \right),
\end{equation}

where $\Sigma_\ell$ is the positive defined covariance matrix (referred to the $\ell$-th attractor) which evolves according to the following time dependent Lyapunov equation:

\begin{equation}
\frac{d}{dt} \Sigma_\ell = J \Sigma_\ell +\Sigma_\ell J^T + D.
\end{equation}

At equilibrium the Fokker-Planck equation converges to a multivariate Gaussian distribution $N(0,\bar{\Sigma}_\ell)$, with covariance given by $J \bar{\Sigma}_\ell +\bar{\Sigma}_\ell J^T + D=0$. 

Summing up, model (\ref{general_model}) can be equipped with $C$  deterministic ($\epsilon=0$) attractors, following the strategy outlined in the preceding Section. These are distinct points in the $N$ dimensional space where the dynamics is embedded.  When accounting for the additional stochastic term ($\epsilon \ne 0$), in the small noise regime ($\epsilon << 1$), the points become multivariate Gaussian clouds, centered around the reference  attractors and with a covariance structure $\bar{\Sigma}_\ell$ that reflects both the deterministic dynamics (via the Jacobian matrix $J$) and the correlation of the imposed noise (through $G$ that enters the definition of $D$). In formulae, the stochastic model can be hence a priori molded to host $C$ stationary stochastic attractors $P_\ell(x)$ with $l=1,...,C$ given by:

\begin{equation}
P_\ell(x) = \frac{1}{\sqrt{\left(2 \pi \right)^N \det(\bar{\Sigma}_\ell) }} 
\exp \left( -\frac{1}{2} (x-\bar{x}^{(\ell)})^T \bar{\Sigma}_\ell^{-1} (x-\bar{x}^{(\ell)}) \right).
\label{asympt_distri}
\end{equation}

Classify, under the stochastic perspective, amounts to steer the noisy dynamics ensuing model (\ref{general_model}) towards distinct asymptotic distributions $P_\ell$, depending on the class of pertinence of the examined items, i.e. the supplied initial condition. Individual elements displaying a degree of intra-class variability will be asymptotically directed towards the very same distribution. Remarkably enough, the dynamical flow triggers the spontaneous emergence of an ordered packaging of data, as reflecting their inherent variability and when the model is integrated in a proper reconstruction pipeline. As we shall prove in the following, this is  the key for a dynamically assisted generation strategy.

\section{The dynamical classifier}
\label{sec:classification}

In this section, we will show that models of the type (\ref{general_model}) can be trained to operate as effective classification tools. Each item belonging to the training set 
acts as an initial condition of the studied dynamical system. The ensuing temporal evolution drives the system towards a stable fixed point (an \emph{attractor}), which flags for the respective class of pertinence. The ability to react differently to the supplied input, depending on the class it belongs, follows the specific form of the coupling matrix $A$. The stochastic term carved via the correlation matrix $G$ enhances the robustness of the scheme. This approach contrasts with traditional feedforward classifiers, where the output is computed in a single forward pass. Here, the decision emerges from the asymptotic state of a dynamical system, with class boundaries implicitly defined by the basins of attraction in state space. As we shall argue, classifying via a genuine dynamical scheme allows the study of convergence time, attractor stability, and robustness to perturbations. To elaborate along these lines, and with no lack of generality, we choose to operate with the Continuous-Variable Firing Rate. We will in particular assume $c=1/8$, $\beta=1$ and $\lambda_\ell=1\ \  \forall\ell$, in what follows. Also $r_i=1$, if not otherwise specified.

Building on the attractor‐planting technique as described in Section
~\ref{subsec:Planting the deterministic attractors}, we link each desired class with one of the system’s attractors. During training, the coupling matrix \(A\) is optimized so that any trajectory initialized at the representation of a training image converges into the basin of the attractor for its corresponding class.

Concretely, given a training set of \(D\) images \(\{x_j\}\), we simulate the dynamics up to a fixed time \(T\gg \tau\), obtaining \(x_j(T)\) for each sample. 
Let $\bar{x}^{(\ell)}$ denote planted attractor for the corresponding class $\ell$-th. The parameters of \(A\) are then updated (actually by acting on the elements of the associated matrix $\tilde{A}$, as specified by equation (\ref{defA})) by minimizing the mean‐squared error
\begin{equation}
L_\ell \;=\; \frac{1}{D}\sum_{j=1}^D \bigl\|x_j(T) - \bar{x}_j^{(\ell)}\bigr\|_2^2,
\end{equation}
ensuring that samples from different classes are driven toward distinct attractors. Classification is then performed by assigning each input to the class label of its final attractor.
Time $T$ should be large enough for the examined dynamical systems to have approximately reached convergence. The training acts self-consistently on the magnitude of the learned elements of $A$ to ensure that this indeed is the case. The convergence loss introduced above works for both deterministic $\epsilon=0$ and stochastic settings $\epsilon \neq 0$. In this latter setting, by requiring that the average trajectory converges to the selected attractors, also insures that the asymptotic distribution is a multivariate distribution of the type (\ref{asympt_distri}), with covariance matrix ${\Sigma}$ given by the associated Lyapunov equation, namely $J \bar{\Sigma} +\bar{\Sigma} J^T + G G^T=0$. 

\begin{figure}[!ht]
    \centering
    \includegraphics[width=0.7\linewidth]{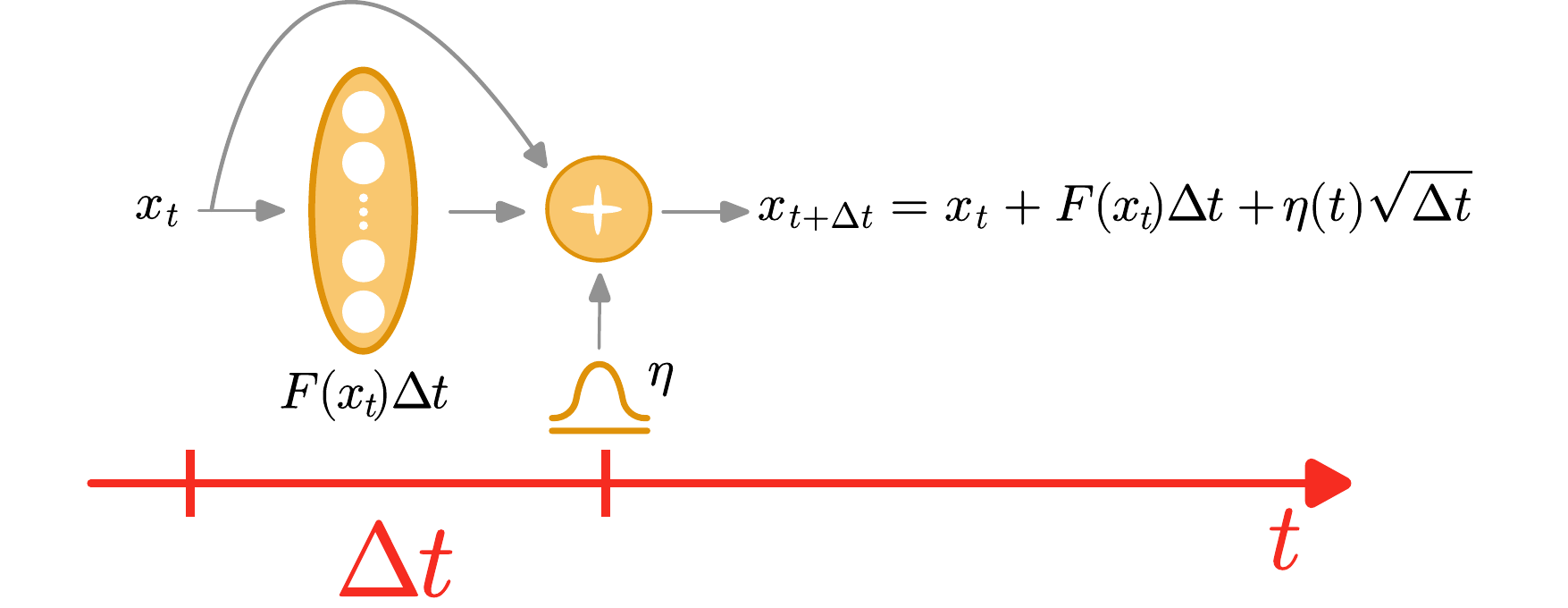}
    \caption{A schematic layout to illustrate the deployment of  model, via its iterative (\ref{general_model}) Euler (or Euler- Maruyama for the stochastic version) representation, on a residual feedforward neural network. Backpropagation through time is performed to propagate the gradients from the output - the late time stage of the dynamical evolution - to the input - the supplied initial condition. }
    \label{fig1}
\end{figure}

 In the deterministic limit ($\epsilon=0$), the evolution of the system is tracked via an Euler's algorithm, deployed on a recurrent feed-forward neural network as illustrated in Figure \ref{fig1}. Identical layers made of $N$ nodes, the size of the model in terms of coupled differential equations, are linked by a linear coupling matrix $A$ where the parameters to be trained are eventually stored. Non linear filters, $f(\cdot)$ and $g(\cdot)$, are punctually applied at the nodes' location, at arrival or departure stages. Time flows along the horizontal axis as depicted  in Figure \ref{fig1}, adjacent layers being separated by a finite amount $\Delta t << 1$. For $\Delta t$ small enough, the trained discrete model behaves like its continuous counterpart. Other integration methods, including Runge-Kutta, can be adopted \cite{rossler2009}, resulting in slightly more involved architectures. The version of the model with $\epsilon \ne 0$ can be handled similarly under the Euler–Maruyama method \cite{higham2001}, customarily employed in Itô calculus to approximate numerical solution of a stochastic differential equation (SDE). For the efficient computation of the gradients through random variables, we employ the celebrated reparametrization trick \cite{kingma2014auto}. In fact, standard optimization tools can now be adopted since the dynamical model to be trained has been {\it de facto} turned into a deep feedforward network made of $T/\Delta t$ nested layers. As a matter of fact, we used the backpropagation through time algorithm to propagate the gradients from the final state towards the input, following a procedure that has been made popular for neural ODEs \cite{chen2018node}. In the example reported hereafter the optimization is carried out by means of the Adam algorithm \cite{kingma2015adam}.

\begin{figure}[!ht]
  \centering
  \setlength{\tabcolsep}{2pt} 
  \renewcommand{\arraystretch}{0} 
  \begin{tabular}{cccccc}
    \includegraphics[width=0.14\linewidth]{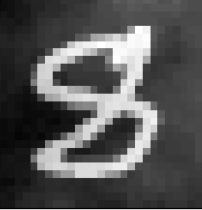} &
    \includegraphics[width=0.145\linewidth]{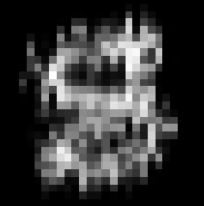} &
    \includegraphics[width=0.14\linewidth]{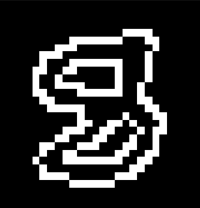} &
    \includegraphics[width=0.14\linewidth]{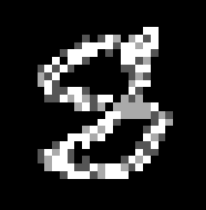} &
    \includegraphics[width=0.145\linewidth]{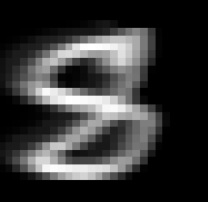} &
    \includegraphics[width=0.14\linewidth]{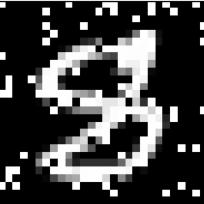} 
    \vspace{0.3cm}
    \\    \small fog & 
    \small glass blur & 
    \small Canny edges & 
    \small Shot Noise & 
    \small motion blur & 
    \small impulse noise
  \end{tabular}
  \caption{Example of corrupted MNIST images from the MNIST-C benchmark \cite{mu2019mnistc}.}
  \label{fig:corrupted-examples}
\end{figure}

As anticipated we will consider the Continuous-Variable Firing Rate as a reference test model. Further, we will challenge the model performance against MNIST, a large database of handwritten digits, ranging from zero to nine, that is commonly used for training various image processing systems and machine learning models. Also, we will make use of a standardized corrupted variant, termed  MNIST-C \cite{mu2019mnistc}, 
a benchmark consisting of 15 image corruptions for measuring out-of-distribution robustness in computer vision. The attractors ($C=10$) are planted following the recipe discussed above, and specifically 
as detailed in formula (\ref{AttrChoice}). A gallery of images from the MNIST-C is displayed in  Figure \ref{fig:corrupted-examples}.

\begin{figure}[!htbp]
    \centering
    \includegraphics[width=1.\linewidth]{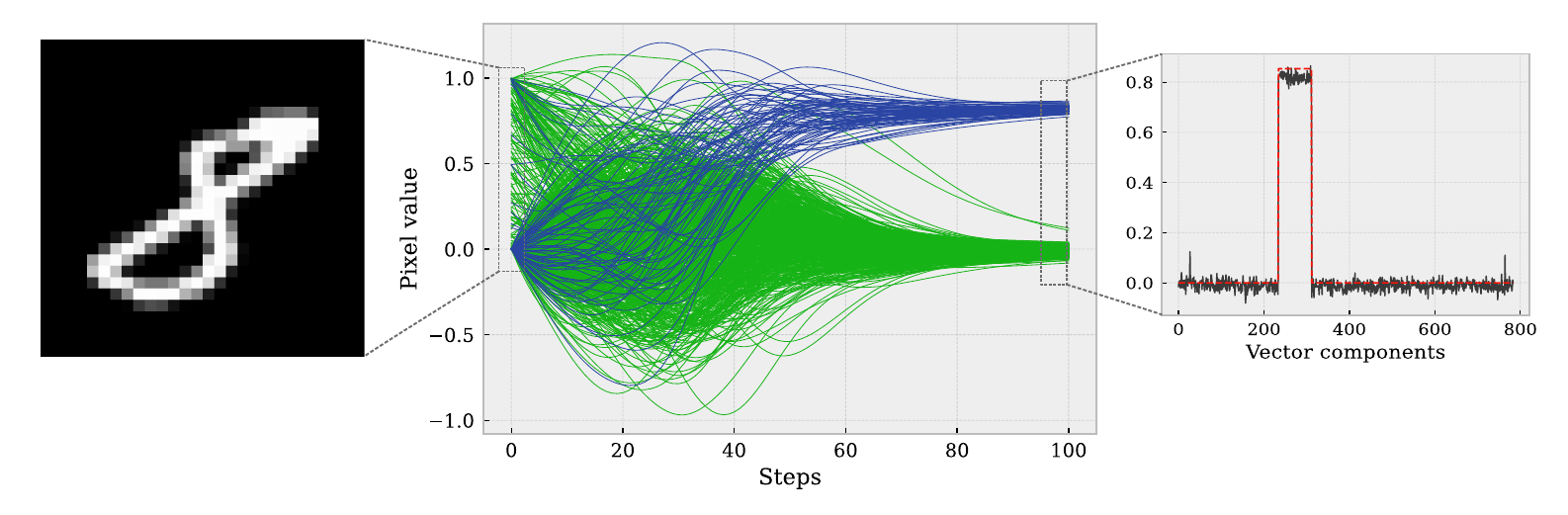}
    \caption{A subset of trajectories, referring to specific entries of the state vector $x(t)$, are plotted against time, for the 
    trained Continuous-Variable Firing Rate with $\epsilon=0$.
    As an initial condition we chose one image belonging to the MNIST test set. The input (here a $28 \times 28$ image of an 8) is unwrapped as a vector of size $784$ and normalized to its largest value. Colors are assigned as reflecting the final value that each pixel is expected to attain, based on the sculpted attractor. Classification is hence correctly performed. }
    \label{fig:evol_1}
\end{figure}

\begin{figure}[!htbp]
    \centering
    \includegraphics[width=1.\linewidth]{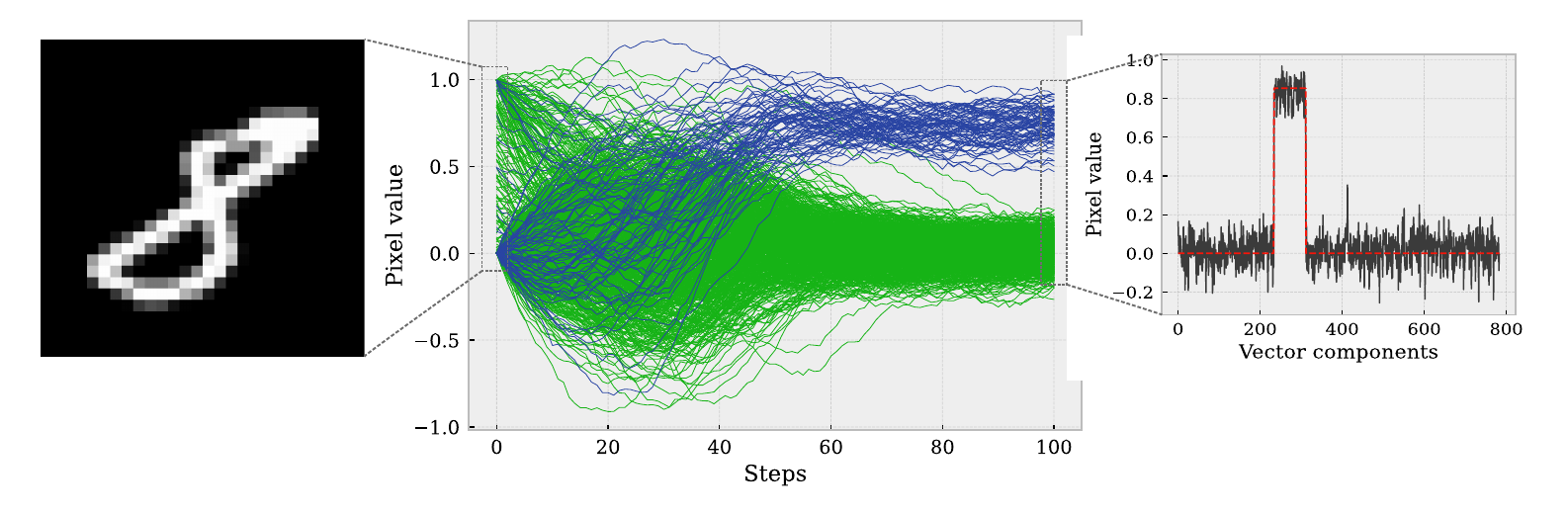}
    \caption{Same as in Figure \ref{fig:evol_1}, but for the stochastic version of the Continuous-Variable Firing Rate. In particular, we set $\epsilon=0.1$ and assume a matrix $G$ to coincide with the identity matrix. We are hence dealing with uncorrelated Gaussian noise, of amplitude $\epsilon$. 
    }.  \label{fig:evol_2}
\end{figure}

In Figure \ref{fig:evol_1} the evolution of the dynamical system is plotted upon training, for one specific supplied example and with reference to the deterministic setting $\epsilon=0$. The image refer to the MNIST dataset and just a subset of the N=784 independent trajectories is plotted to help visualization. Following the injected initial condition - one digit belonging to the test set - the system evolves for a transient order $T$ out of equilibrium before eventually landing on the asymptotic attractor assigned to the class of reference. Different colors reflect the final value that pixels are expected to attain, depending on the specificity of the shaped attractor. All reported trajectories head towards the correct destination target thus implying that the input gets duly classified. In Figure \ref{fig:evol_2} the same analysis is reported for the stochastic trained model. Trajectories, depicted with an analogous color code as illustrated above, are fluctuating in time thus  
reflecting the stochastic nature of the model. Interestingly enough the imposed noise source is kept active at late time. Here, the dynamical system samples noisy attractor that organizes around the designated deterministic fixed point, marking a clear distinction with past attempts to operate classification via  stochastic ODEs \cite{chicchi2025deterministic}. In this latter setting, in fact, the strength of the noise is tuned by ad hoc multiplicative factor, engineered to fade away when the system approaches any of the prescribed attractors. As we will discuss in the following, the convergence towards a stationary distribution (which is here structurally enforced by the mathematical formulation)  is a key property to turn the dynamical model into a fully generative scheme. In this initial example, we freeze the correlation noise matrix $G$ to the identity and solely train the elements of the coupling adjacency matrix $A$ for the model to learn the classification task.   

To further elaborate on the role played by noise, we record the performance of the trained models against the aforementioned databases (MNIST and MNIST-C) for three different settings:

\begin{enumerate}
    \item We train the deterministic model ($\epsilon=0$) against the pristine MNIST database, and test the performance of the generated model against the test set for both MNIST and MNIST-C. Training amounts to optimize the elements of the adjacency matrix $A$.    
    \item We train the stochastic model for different choices of $\epsilon \ne 0$ and $G \equiv  \mathbb{I}$ against MNIST and evaluate the ensuing performance against both MNIST and MNIST-C. Also in this case the optimization acts on matrix $A$. 
    \item We train the stochastic model against MNIST with both  $A$ and $G$ trainable. The performance are evaluated against both MNIST and MNIST-C. In this latter setting we introduce one further term in the loss function that prevents the model from converging to the trivial solution $G_{ij}=0$, $\forall i,j$. More specifically, the loss is equipped with the additional factor $\alpha/(\sum_{i,j} |G_{ij}|)$, where $\alpha$ is an hyperparameter of the model.   
    
\end{enumerate}

\begin{figure}[!htbp]
    \centering
    \includegraphics[width=0.8\linewidth]{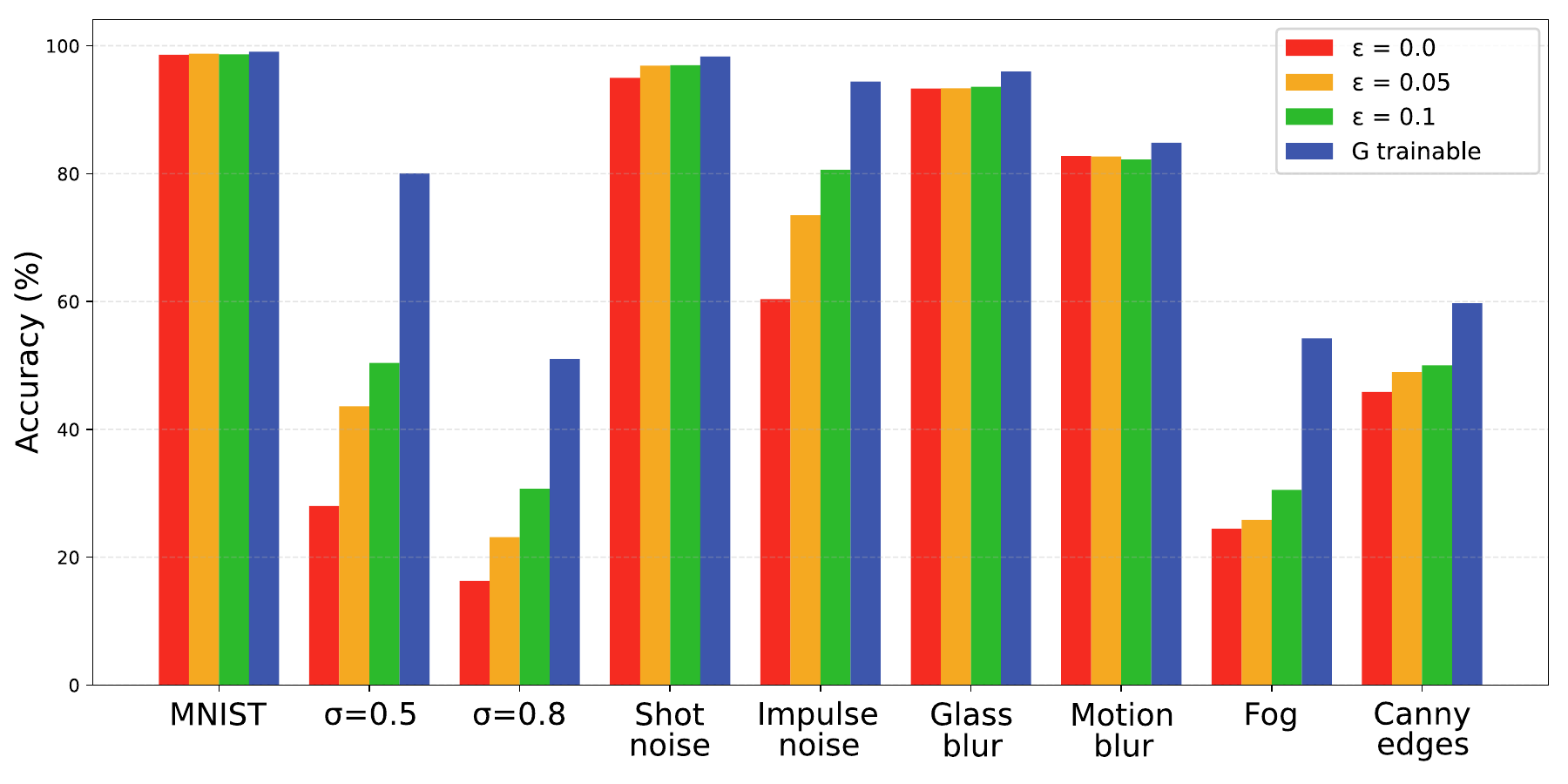}
    \caption{Visual representation of the accuracy of different models ($\epsilon=0$, $\epsilon=0.05$ and $G= \mathbb{I}$, $\epsilon=0.1$ and $G= \mathbb{I}$, $\epsilon=1$ and $G$ fully trainable) trained on the uncorrupted MNIST model and tested against data gathered from MNIST-C. The reported accuracies are specifically referred to data perturbed with Gaussian noise with $\sigma \in \{0.5, 0.8\}$, shot noise, impulse noise, glass blur, motion blur, fog, Canny edges.}
    \label{fig:accuracy_corrupted_dataset}
\end{figure}

The results of the analysis are visually depicted in Figure     \ref{fig:accuracy_corrupted_dataset} and display a clear trend:
moderate noise injection during training acts as an effective regularizer of the trained dynamics, improving robustness to distribution shifts without degrading clean-data performance. As an example, under Gaussian noise with amplitude $\sigma=0.8$, the accuracy rises from $16.30\%$ (for a deterministic model trained on uncorrupted MNIST data) to $51.00\%$ (for a model that seeks to optimize both $A$ and $G$ when trained on the unperturbed MNIST). The model therefore benefits from adapting the noise correlation to the analyzed data, in that it acquires an additional degree of inherent flexibility that proves useful when handling corrupted data during deployment.

The performances displayed in Figure \ref{fig:accuracy_corrupted_dataset} are made explicit in Table   \ref{table:accuracy}, where the reported figures (mean and standard deviation) are computed over multiple runs.

\begin{table}[!htbp]
\centering
\resizebox{\textwidth}{!}{%
\begin{tabular}{lccccccccc}
\hline
\textbf{Model} & \textbf{MNIST} & $\bm{\sigma=0.5}$ & $\bm{\sigma=0.8}$ & \textbf{Shot noise} & \textbf{Impulse noise} & \textbf{Glass blur} & \textbf{Motion blur} & \textbf{Fog} & \textbf{Canny edges} \\
\hline
$\epsilon \!=\! 0.0$  & 98.58 $\!\pm\!$ 0.30 & 28.01 $\!\pm\!$ 0.23 & 16.30 $\!\pm\!$ 0.27 & 94.97 $\!\pm\!$ 0.23 & 60.38 $\!\pm\!$ 0.25 & 93.32 $\!\pm\!$ 0.24 & 82.77 $\!\pm\!$ 0.21 & 24.47 $\!\pm\!$ 0.24 & 45.87 $\!\pm\!$ 0.22 \\
$\epsilon \!=\!0.05$ & 98.76 $\!\pm\!$ 0.28 & 43.60 $\!\pm\!$ 0.23 & 23.14 $\!\pm\!$ 0.29 & 96.89 $\!\pm\!$ 0.25 & 73.50 $\!\pm\!$ 0.31 & 93.35 $\!\pm\!$ 0.26 & 82.69 $\!\pm\!$ 0.30 & 25.82 $\!\pm\!$ 0.31 & 48.96 $\!\pm\!$ 0.26 \\
$\epsilon\!=\!0.1$  & 98.68 $\!\pm\!$ 0.26 & 50.40 $\!\pm\!$ 0.23 & 30.70 $\!\pm\!$ 0.24 & 96.97 $\!\pm\!$ 0.23 & 80.62 $\!\pm\!$ 0.32 & 93.59 $\!\pm\!$ 0.22 & 82.22 $\!\pm\!$ 0.46 & 30.52 $\!\pm\!$ 0.25 & 50.01 $\!\pm\!$ 0.31 \\
G train. & \textbf{99.08 $\!\pm\!$ 0.06} & \textbf{80.02 $\!\pm\!$ 0.08} & \textbf{51.00 $\!\pm\!$ 0.09} & \textbf{98.32 $\!\pm\!$ 0.13} & \textbf{94.41 $\!\pm\!$ 0.12} & \textbf{95.98 $\!\pm\!$ 0.17} & \textbf{84.84 $\!\pm\!$ 0.17} & \textbf{54.24 $\!\pm\!$ 0.20} & \textbf{59.72 $\!\pm\!$ 0.19} \\
\hline
\end{tabular}%
}
\caption{Comparison of classification accuracies (\%) for different models on on MNIST and corrupted variants.}
\label{table:accuracy}
\end{table}

From a dynamical–systems perspective, noise perturbs trajectories within the reference basins of attraction, by forcing the classifier to learn decision boundaries that remain stable under the imposed perturbations. The model trained on immaculate data proves therefore effective also when tried in contexts where adversarial attacks manifest as possible source of corruption. Importantly, the observed gain is achieved without sacrificing clean-data performance.  

Recall that the classification is operated upon measuring the distance between the average asymptotic trajectory - as triggered by the supplied initial condition - and the positions of the planted attractors. The assigned class follows by identifying the closest possible target. Computing the late time average from the dynamical trajectory, is equivalent to accessing the deterministic approximation of the generated stochastic signal. Hence, the performances of any trained stochastic model match exactly those
obtained when when deploying a deterministic analogue ($\epsilon=0$) with an identical matrix $A$. Training under the stochastic perspective, leaves therefore a non trivial imprint in the optimized adjacency matrix $A$. This yields a more effective deterministic classifier (in terms of  robustness against adversarial random attacks, of the type outlined above) as compared to what one can eventually get under a fully deterministic angle.  

We recall that in the proximity of a planted attractors the Lyapunov equation arises naturally as a viable tool to quantify the covariance of the expected distribution, under the linear noise approximation, thus assuming small to moderate noise strengths. This latter is a priori set by the set hyper-parameter $\epsilon$, when dealing with $G=\mathbb{I}$, or it can be assessed a posteriori by measuring the norm of the optimized matrix $G$. In all the examples here examined the system converges to an optimal matrix $G$ with a sufficiently small norm for the linear noise approximation to hold true. The stationary covariance $\bar{\Sigma}_\ell$ contains information on  the shape of the stochastic attractors, which are automatically molded by the self-consistent optimization problem, when $G$ is let free to vary. Isotropic $\bar{\Sigma}_\ell$ implies uniform robustness in all input directions, whereas anisotropic $\bar{\Sigma}_\ell$ indicates that certain directions tolerate more variation than others.  
Estimating $\hat{\Sigma}_\ell$ from late-time trajectories allows us to connect noise parameters, as encoded in $G$, to the empirical robustness under corruptions. In particular, given $\Sigma_{\ell}$ and $\Sigma_m$, the Mahalanobis $\Delta_{\ell m}$ separation between attractors $\bar{x}^{(\ell)}$ and $x^{(m)}$ can be computed as

\begin{equation}
\Delta_{\ell m} = \big(\bar{x}^{(\ell)}- \bar{x}^{(m)}\big)^T \Sigma_\ell^{-1} \big(\bar{x}^{(\ell)} - \bar{x}^{(m)}\big).
\end{equation}

Large $\Delta_{\ell m}$ values correspond to well-separated distribution and, thus local basins of attraction with minimal mutual interference. For the case at hand, when matrix $G$ is being adjusted all along the training, and for MNIST as the testbed reference dataset, we observe consistently larger Mahalanobis distances between the trained stochastic attractors. This indicates that the attractors are more clearly separated in a statistical sense, thereby reducing overlap among class-conditional distributions. Such an effect provides a first explanation of why introducing a trainable noise covariance matrix can improve classification accuracy, as it facilitates more distinct and reliable attractor dynamics. For further quantitative details refer to the annexed Appendix \ref{App2}.

\begin{figure}[htbp]
    \centering
    \subfloat[Deterministic]{%
        \includegraphics[width=0.48\linewidth]{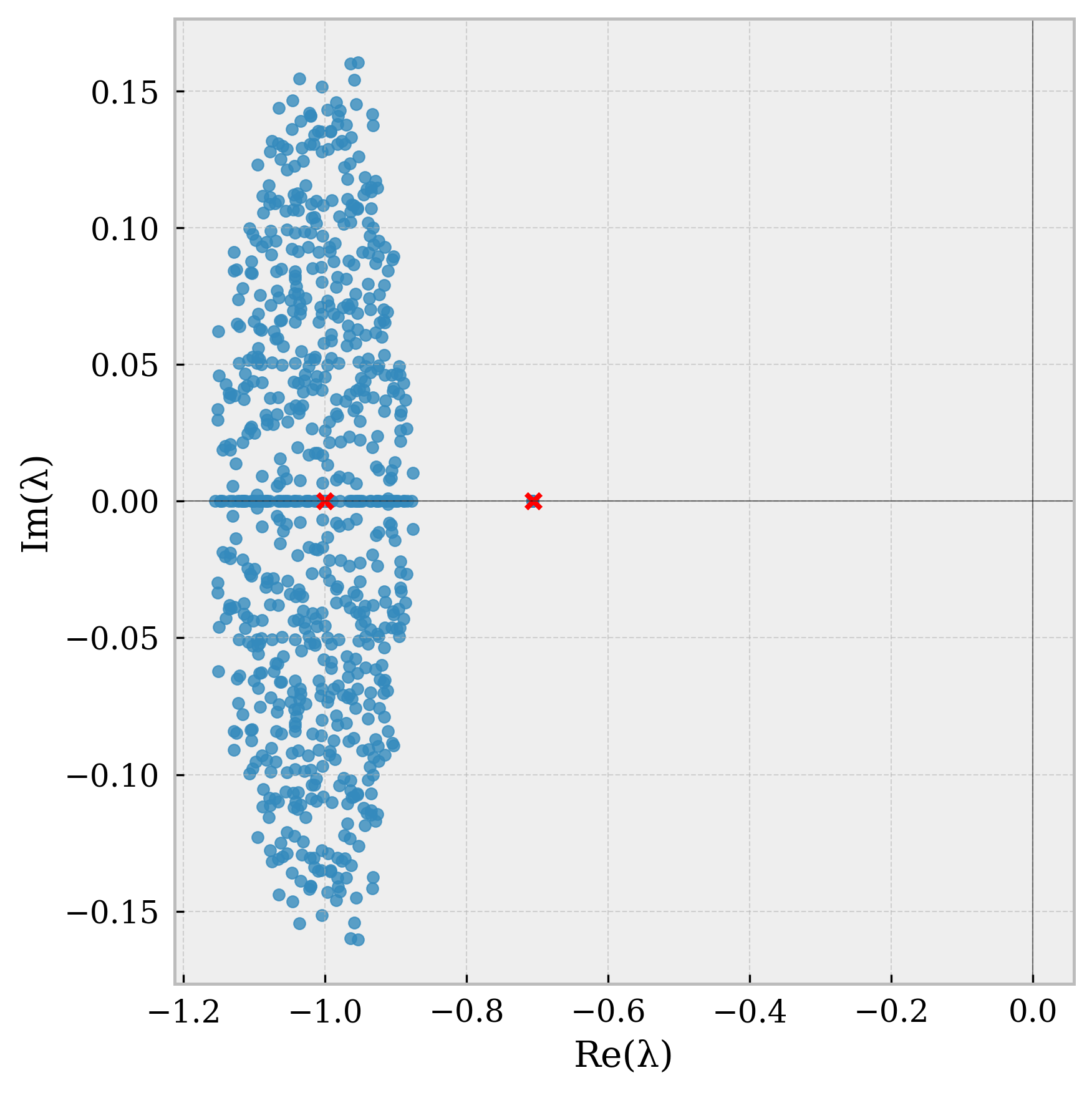}%
        \label{fig:image1}%
    }\hfill
    
    \subfloat[Stochastic]{%
        \includegraphics[width=0.48\linewidth]{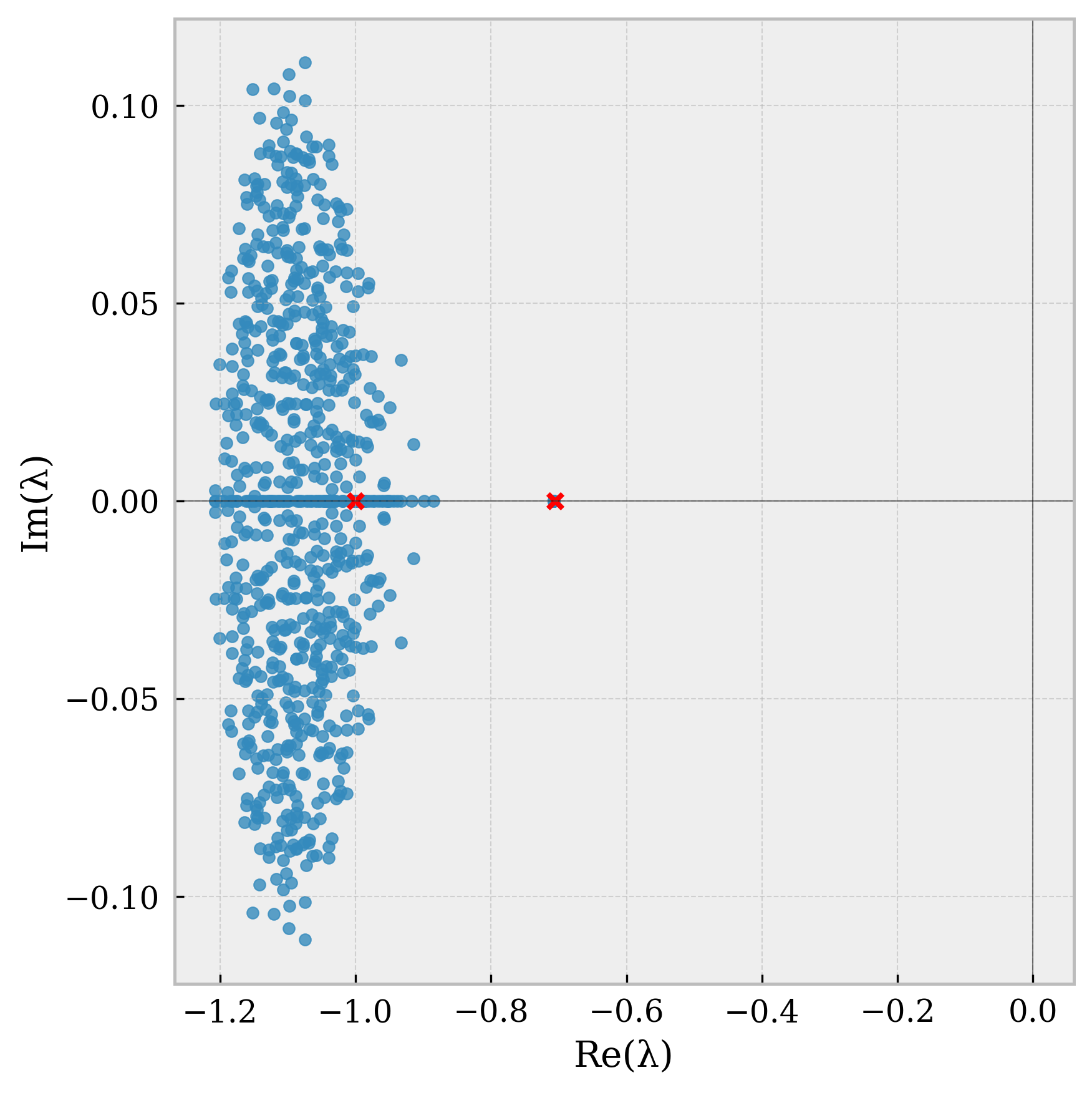}%
        \label{fig:image2}%
    }

    \caption{Spectrum of the Jacobian matrix computed in 
     correspondence of one of the planted attractors for both the 
     deterministic (panel a) and stochastic (panel b) versions of the Continuous-Variable Firing Rate model trained against MNIST data. Blue points to the left indicate stable modes; points further left correspond to faster convergence rates to the equilibrium attractor. Nonzero imaginary parts $\mathrm{Im}(\lambda)$ signal oscillatory convergence. Note that every depicted spectra share a set of common eigenvalues (as identified by the red crosses), in agreement with what explained in Appendix \ref{App1}.}
    \label{fig:Jacobianstochasticcomparison}
\end{figure}

The stability of the planted attractors is a byproduct of the training. The eigenvalues of the Jacobian matrix associated to each of the planted target attractors  consistently spread in the negative portion of the complex plane, implying that stability has been achieved.  Remarkably, and this is consistently observed throughout the experiments, the presence of noise makes the system slightly more stable: the cloud of the Jacobian's eigenvalues drifts towards the left, within the complex plane, as compared to the what it is found, under identical operating conditions, for the corresponding deterministic scenario.  To report on this issue, we focus on both  deterministic and stochastic versions of the Continuous-Variable Firing Rate model trained against MNIST data, and plot in Figure \ref{fig:Jacobianstochasticcomparison} the computed Jacobian eigenvalues calculated in correspondence of one of the planted attractors.  

Beside this specific aspect, the spectrum of the Jacobian linearized around each class attractor provides a practical - a posteriori - diagnostic for assessing whether the model has learned stable and well-conditioned dynamics.   The spectral perspective reveals not only the overall stability (sign of the real parts of the eigenvalues) but also the distribution of convergence rates across modes, the presence of oscillatory behaviour through complex conjugate pairs, and the existence of slow or marginally stable directions that may affect robustness. Unlike a single stability coefficient, the full eigenvalue cloud provides a richer, mode-by-mode picture of how trajectories contract toward the attractor after training.

\section{The generation scheme}
\label{sec:generation}

In this section, we investigate the Continuous-Variable Firing Rate model (as a mere representative example of the general class of models here considered) as a generative scheme. We will in particular focus on different tasks of image generation.  A unified framework will be proposed which can integrate both classification and generation. Within this perspective, we also explore the potential of the method for conditional generation. We will begin by considering a setting where the Continuous-Variable Firing Rate model is solely employed in the compression phase, from the scrutinized data to the distribution in the latent space. The convergence to a latent distribution, centered around a well defined mean for each of the classes of interest, is a property that gets naturally inherited from the dynamical nature of the classifier, as introduced above. In the second part of the reported analysis, a stochastic dynamical model is introduced in the pipeline to sample from the latent space to invert the process, back to the data distribution. As it will become transparent from the above, the proposed setting share some degree of architectural similarity with Variational Autoencoders (VAE) \cite{kingma2014auto} a type of deep learning model that learns a compressed, probabilistic latent space representation of data, enabling them to generate new data similar to the training set. Following our proposed approach, the compression and decoding phases can be operated with effective stochastic dynamical model. On the one side, this enables for the probabilistic latent representation to emerge naturally as a byproduct of the formulated dynamical scheme. On the other, we provide a concrete example of how sensible models, relevant to computational neuroscience, could be effectively used to assemble  generative tools with notable performance. In diffusion-based neural networks \cite{10.1145/3626235} samples are evolved to eventually populate a normal distribution that provided the latent representation of the examined data. By sampling from this latter distribution and reversing the (dynamical) diffusion process yields original high quality images. This process requires however approximating the non linear score function with an implicit deep neural networks. This is at variance with the proposed setting, where explicit dynamical modules can be employed (at least in principle, see final application) both in compression and generation modes.
\subsection{The dynamical model bridges the gap from the input to the latent representation}

In the first of the aforementioned schemes, the following approach is adopted. First, images are compressed into a latent representation through an Encoder: $X \xrightarrow{~E~} x(0) \in \mathbb{R}^d $. Once in latent space, the state evolves under the Continuous-Variable Firing Rate dynamics with planted attractors, integrated via the Euler–Maruyama method. Asymptotically, the system is bound to converge towards a Gaussian distribution centered at the corresponding attractor $\bar{x}^{l}$ and with covariance $\Sigma_\ell$ given by the associated Lyapunov equation. At finite time $T$, the initial distribution is turned into an approximately Gaussian, still centered at $\bar{x}^{(\ell)}$, with an empirical covariance $\bar{\Sigma}_\ell$, which can be computed numerically. This is the latent space, that we will be sampling for reconstruction (and generation) purposes as in the spirit of a VAE philosophy. At t $\rightarrow \infty$, the system has lost track of the information as stemming from the initial distribution of the analyzed dataset: images of distinct initial points are indeed populating the target Gaussian, with no inherent organization that reflects the specificity of the provided input. At variance, at time $T$, the output of the dynamical systems yields a compact distribution which can be ideally fragmented into a collection of contiguous blobs, each one associated to a peculiar subgroup of coherent input elements. Input elements which appear to share common traits, as categorized by the dynamical update rule, are packed closely within a bound latent space centered around the attractor. Stated differently, the dynamics implements an ingrained manifold disentanglement, with no need for an ad hoc twinkling of the relevant parameters. The unsupervised discovery of disentangled latent factors is thus inborn into the convergent stochastic dynamics. 

The final state $x(T)$, namely the produced image in latent space, is then passed to a decoder $D$, implemented via a standard feedforward neural network. This latter can be replaced by yet another dynamical unit, as we will comment in the following. In short, the full process can be represented by the following scheme: 
\[
X \xrightarrow{~E~} x(0) \in \mathbb{R}^d 
\;\xrightarrow{~\text{Stochastic CVFR to } T~}\;
x(T)
\;\xrightarrow{~D~}\;
\widehat X,
\]

where $E$ stand for the encoder which maps into a space of dimension $d$. The forward evolution is implemented via the Euler–Maruyama update scheme, as discussed with reference to the classification mode of operation. To enforce pixel-wise fidelity between the input image and its reconstruction, we postulate a reconstruction loss of the type introduced below:
\[
\mathcal{L}_{\text{rec}} = \frac{1}{ND}\sum_{i=1}^N \big\| X_i - \widehat{X}_i \big\|^2,
\]
where \(N\) is the number of training images, \(D\) the input dimensionality, \(x_i\) the original image, and \(\widehat{x}_i\) its reconstruction. Further we, require  
convergence to the imposed attractor(s), with the following loss term: 
\[
\mathcal{L}_{\text{conv}} = \frac{1}{Nd}\sum_{i=1}^N \big\| x_i(T) - x^{(\ell=y_i)} \big\|^2,
\]
where \(d\) is the latent dimension and \(\bar{x}^{(\ell)}\) is the attractor associated with class \(\ell\). This forces latent trajectories to approach the intended attractor. This mechanism also enables conditional generation, since multiple attractors can be planted.  Finally we introduce an additional loss term to preserve a symmetric distribution of data around each attractor:

\begin{equation}
    \mathcal{L}_{\text{centroid}} = \frac{1}{dC}\sum_{\ell=1}^C \left\| \frac{1}{N_\ell} \sum_{i=1}^N \delta_{y_i,\ell}\,x_i(T) - \bar{x}^{(\ell)} \right\|^2,
\end{equation}
where \(C\) is the number of classes and \(N_\ell\) the number of samples in class \(\ell\).  In the literature this is referred to as to a centroid-consistency term. 

\begin{figure}[!htbp]
    \centering
    \includegraphics[width=0.7\linewidth]{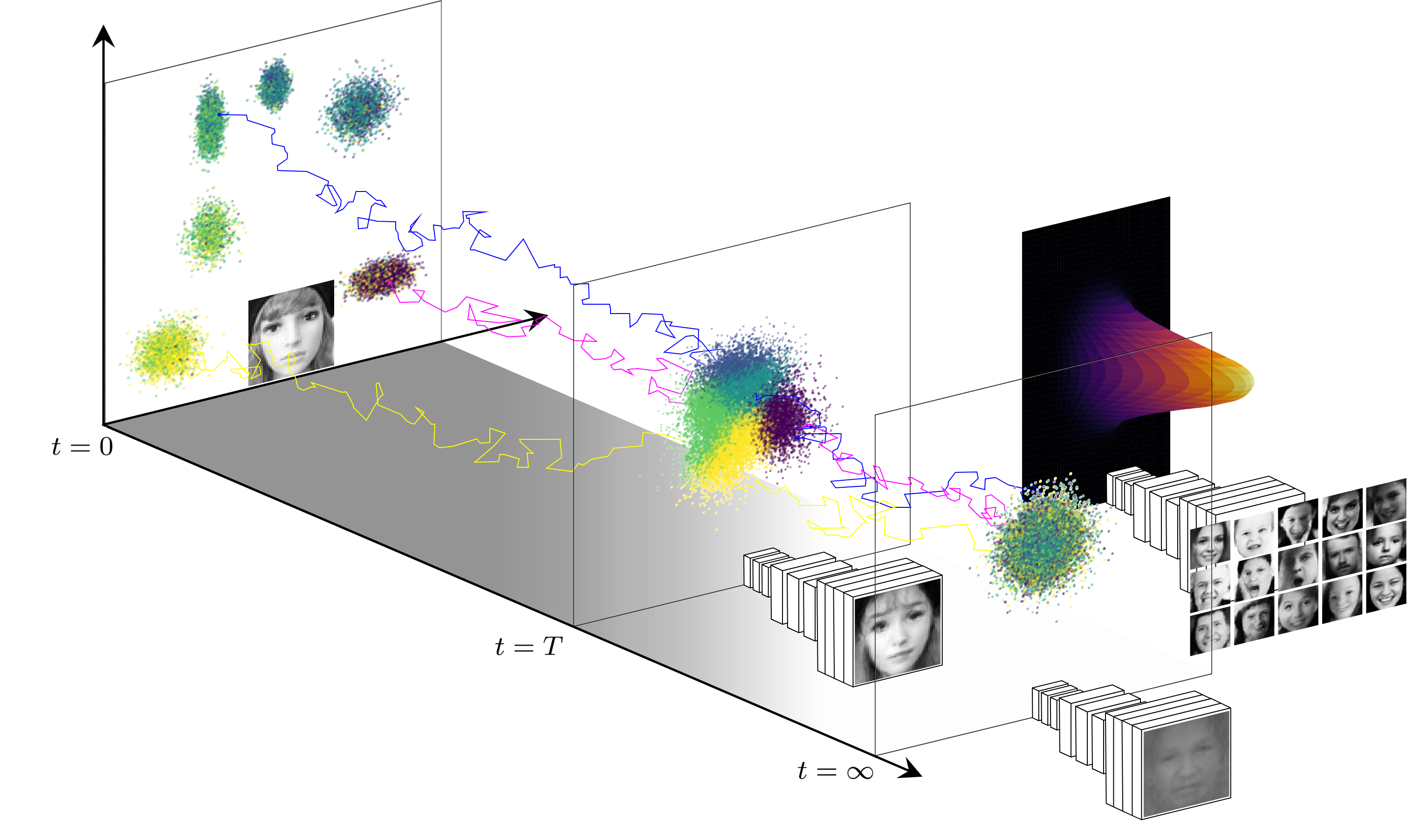}
    \caption{A pictorial representation of the generative scheme as outlined in the main body of the paper. The supplied input (here an image of a face) is provided as n input to the trained dynamical model. Asymptotically the system is bound to converge towards a multivariate Gaussian distribution centered around the target distribution. At time $T$ (to be chosen freely, as the dynamical system self-adjusts, along train, to cope with the choice made), data are packed in adjacent 'bubbles' that fill a compact domain in latent space. Sampling the empirical distribution at time $T$ (a Gaussian distribution, centered at the $\ell$-th attractor(s) and with empirical covariance matrix $\bar{\Sigma}_\ell$) and feeding a standard feedforward module yields a scheme with full generative power. This latter feedforward unit can be replaced by yet another dynamical model of the type here considered. Hence, both compression to the latent space and the subsequent generation scheme can be traced back to fully transparent dynamical units.}
    \label{fig:scheme}
\end{figure}

Reconstruction and convergence losses mutually balance: on the one side the convergence to the attractor forces the trajectories to stay as close as possible to the target stationary solution. On the other, 
for the reconstruction to work properly, trajectories should be prevented to collapse on just one point of the embedding space: hence, the reconstruction loss keep trajectories as separated as possible. Subject to this contrasting coercing effects, the dynamical model self-adjusts automatically to have the distribution in the latent space aligned with the asymptotic one (recall that the specific form of the final attractor is embedded in the dynamics, under the linear noise ansatz and via the associated Lyapunov equation), while preventing  single image probability to merge one another. Intuition tells that as trajectories approach the deputed attractor, their probability 'bubbles' begin to overlap, but not too much due to the reconstruction term. Decoding from these mixed neighborhoods can yield \emph{novel} samples that were not present in the training set, distinguishing generation from mere memorization. 
Noise \(\epsilon\) is therefore not only a regularizer: it controls overlap and diversity in the generation process. In Figure 
\ref{fig:scheme} a pictorial representation of the general framework, as qualitatively outlined above,  is shown. In the following, we will provide quantitative evidences to substantiate the above description.  

Summing up, the model is equipped with a global loss $\mathcal{L}$, which can be cast in the form: 

\begin{equation}
\scalebox{0.8}{$
    \mathcal{L} = \underbrace{\frac{1}{Nd}\sum_{i=1}^N||x_i(T)-x^{(\ell=y_i)}||^2}_{\text{attractor convergence}}
     + 
    \underbrace{
    \frac{1}{ND}\sum_{i=1}^N||x_i-\widehat{x_i}||^2}_{\text{reconstruction}} +
    \underbrace{
    \frac{1}{dC}\sum_{\ell=1}^C\left|\left|\frac{1}{N_\ell}\sum_{i=1}^N\delta_{y_i,\ell}x_i(T) - x^{(\ell=y_i)} \right|\right|^2}_{\text{centroid consistency}},
    $}
\end{equation}
where $\{x_1, \dots, x_N\}$ is the training dataset, $\widehat{x}$ is the reconstructed image, $x(T)$ is the state of the system at time $T$, $x^{(\ell=y)}$ is the attractor corresponding to the label $y$ and $N_\ell=\sum_{i=1}^N\delta_{y_i,\ell}$, namely the number of elements of a given class. As already argued, the first term favors full convergence, while the second advocates for a sort of structural significance for an adequate reconstruction of the original data. The third term forces data to fill the space around the attractor.

When it comes to the training, it should be pointed out that all the blocks are trained at the same time. During training, one can clearly choose to deal with more than just one attractor. The model turns thus into a conditional generative scheme. Indeed, the implemented loss allow the model to separate different classes of images in different basin of attraction, by shaping the ensuing dynamics of the system. Once the training is completed, by sampling from different basin one can access to generation of different classes of images, with no need for training an had hoc discriminator, as its is usually done under customary generative frameworks. One can otherwise decide to plant just one attractor and, in this case, one falls back in the classic generative process in which all the images are generated from the the same latent distribution. In the experiments that we will report here below the training acts on the free parameters of the feedforward module (both the encoder and decoder) and on the entries of the adjacency matrix $A$. Matrix $G$ that defines the covariance of noise is set to the identity matrix. The strength of the noise is controlled by the parameter $\epsilon$, which will be tuned at will to quantitatively elaborate on the role played by the imposed stochastic drive. Once the training process is ended, generation of new data for class $\ell$ is achieved by sampling from a multivariate normal distribution $\mathcal{N}(\bar{x}^{(\ell)},\Sigma^{(\ell)}_T)$ where $\Sigma^{(\ell)}_T$ is the empirical covariance estimator
\begin{equation}
    \Sigma^{(\ell)}_T=\frac{1}{N_{\ell}-1}\sum_{i=1}^{N_{\ell}}\left[x_i(T)-\bar{x}^{(\ell)}\right]\left[x_i(T)-\bar{x}^{(\ell)}\right]^T
\end{equation}
evaluated on the sub-dataset of the elements belonging to the class $\ell$. The samples are then fed to the expansive neural networks, that translates the latent variables into new genuine images. 

\begin{figure}[!htbp]
    \centering
    \includegraphics[width=1.\linewidth]{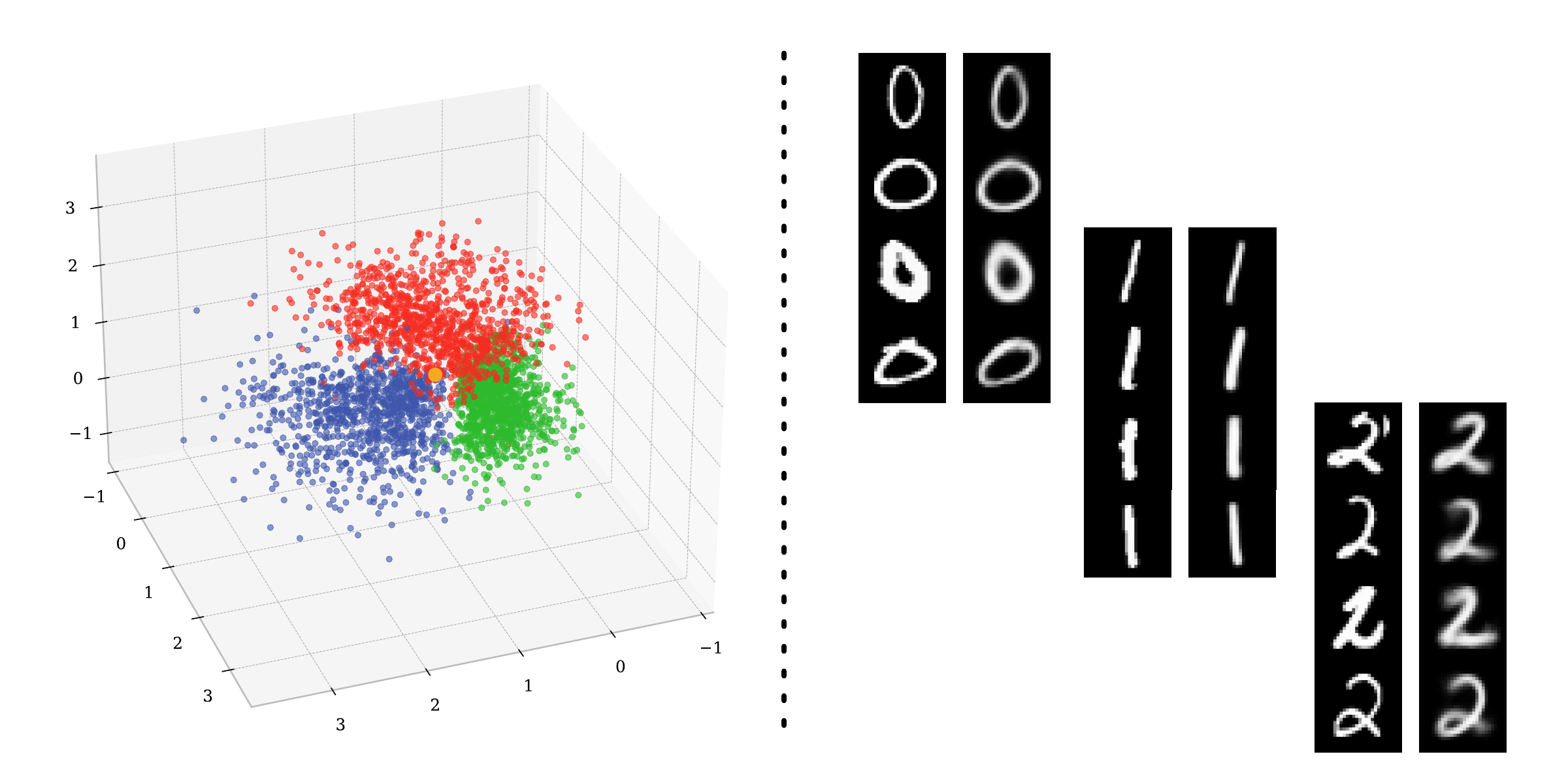}
    \caption{To shed light on the inherent ability of the dynamical system to favor a spontaneous clusterization in latent space, we consider a  3-dimensional representation of the evolved data distribution around the attractor. This is achieved by forcing the decoder to process with just the first three components of the latent dimension as an input. Right panel: a few reconstructed digits are compared to their original homologue. The quality of the reconstruction is obviously affected by the choice of relying on just three scalar inputs for the reconstruction pipeline. Left panel: the first three elements of the obtained latent vectors (the  output $x(T)$),  namely those used for reconstruction purposes, are plotted with a choice of the color that reflects the class to which the evolved data belongs to. The data populated contiguous thought visually separated blobs, which bare distributed around the target attractor (yellow point). Here, the dynamical system is evolved for 100 steps of duration $\Delta t = 0.03$, hence $T=3$, in arbitrary units.}
    \label{fig:Clusters}
\end{figure}

Let us being by reporting a preliminary investigation to shed light onto the distributions of the dynamically evolved input data-points at time $T$. To this end we work with the MNIST database restricted to just three digits, and specifically numbers 0,1 and 2. The operated setup includes a simple feed-forward encoding module that reduces the data dimension from $D=784$ to $d=20$. Evolution is then carried out towards the latent space by integrating the Continuous-Variable Firing Rate model, deployed on a recurrent feedforward scheme, following the prescription of the Euler-Maruyama algorithm. The reparametrization trick is used to handle the stochastic components. A meaningful three-dimensional representation of the evolved  distribution is obtained by forcing the decoder to get as an input only the first three components of the latent vector. The quality of the reconstruction can be appreciated by visual inspection of Figure \ref{fig:Clusters}, see left panel. In the right panel of the same Figure, the three entries of the latent vectors used for reconstruction purposes are plotted with a color code that reflect the specific class of pertinence of the input item. As anticipated on the basis of an intuitive reasoning, the data organize in separated cluster, homogeneous in colors, distributed around the target attractor, here depicted with a yellow point.

\begin{figure}
    \centering
    \includegraphics[width=1.\linewidth]{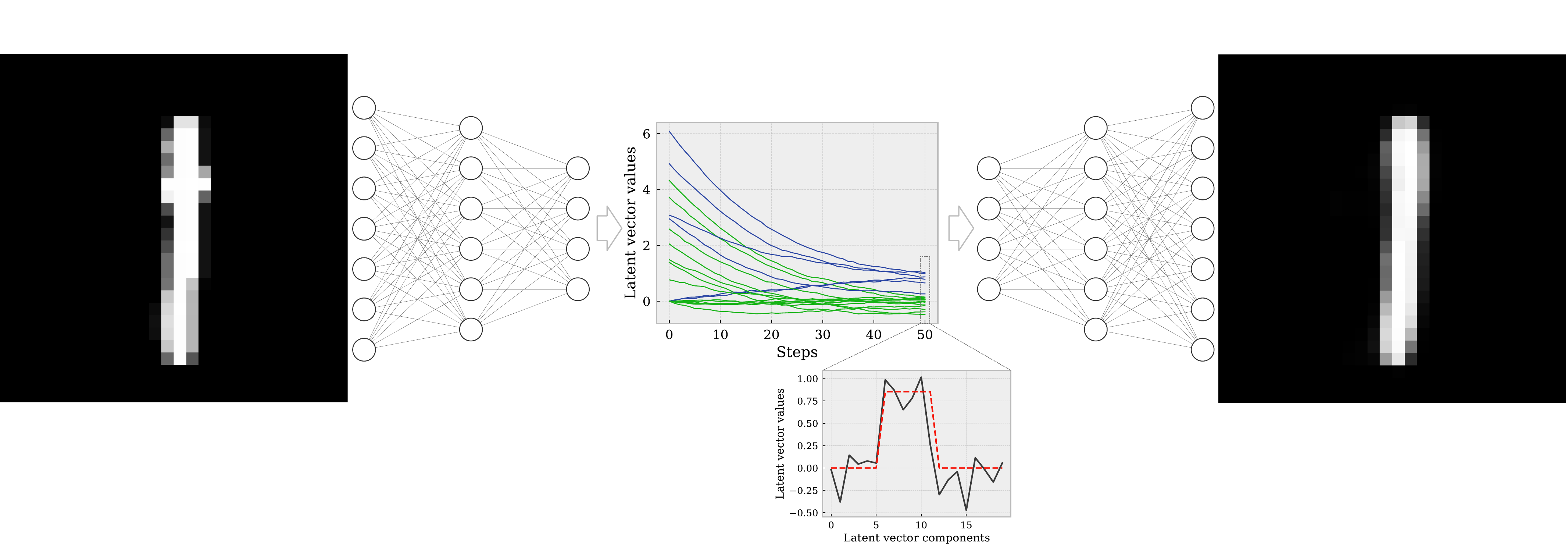}
    \caption{An image representing a number one is processed by the encoded  and provided as an input to the Continuous-Variable Firing Rate model. The trajectories are colored (blue vs. green) depending on the specific destination target that they are bound to reach for a correct classification of the supplied input. In the inset displayed in the lower portion of the Figure, the asymptotic attractor (red dashed line) is compared with the average trajectory (black solid line). The nice correspondence between the two reported curves implies that the correct attractor has been eventually hit.  The decoder takes the produced dynamical output and  transforms it into a reconstructed version of the supplied input digit.}
    \label{rec_minst}
\end{figure}

By relaxing the constraints on the dimensions to be used for generation, the quality of the reconstruction improves significantly.
Figure \ref{rec_minst} provides a pictorial representation of the whole process as described above. Here we leverage on the whole latent space for size $20$. The supplied digit,  one individual representative of the reservoir of ones, is passed through the encoding step and fed as an input to the dynamical system. The ensuing evolution (black solid line) makes the system to approach - on average - the asymptotic attractor (red dashed line), as shown in the annexed panel. Notice also that each pixel heads towards the corresponding destination target, as the segregation in colors (blue vs. green) of the resulting trajectories clearly exemplify. The decoder transforms the dynamical output into a reconstructed version of the input digit. The decoder can be then operated as an image generator. To this end, different data points sampled from the latent distribution (as described above) are provided as an input to the feedforward decoder. 

\begin{figure}[!htbp]
    \centering
    \includegraphics[width=0.8\linewidth]{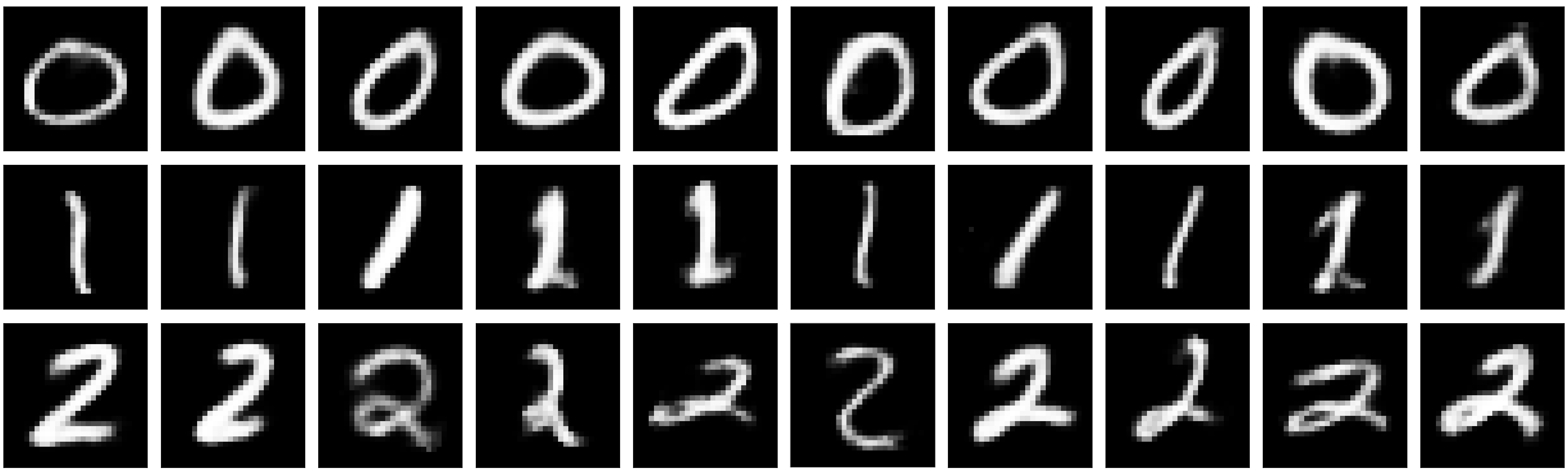}
    \caption{Conditional generation of MNIST images. Three different attractors are considered, one for each of the classes being examined. Generated images are obtained by sampling a latent space made of $20$ independent entries (the actual size of the evolved dynamical system, after encoding).}
    \label{fig:generationMNIST}
\end{figure}

In Figure \ref{fig:generationMNIST} we report a few example of generated digits. It should be stressed that we have here implemented a conditional generation, meaning that different numbers are associated to distinct attractors. These latter are engineered by following the recipe describe above. Remarkable is in particular the repeated occurrence of digits one with an horizontal trait underneath the vertical bar. This way of crafting the ones is very rare across the original MNIST dataset and probably results from a non trivial merging of digits one and two, as packed in latent space.

\begin{figure}[!htbp]
    \centering
    \includegraphics[width=0.5\linewidth]{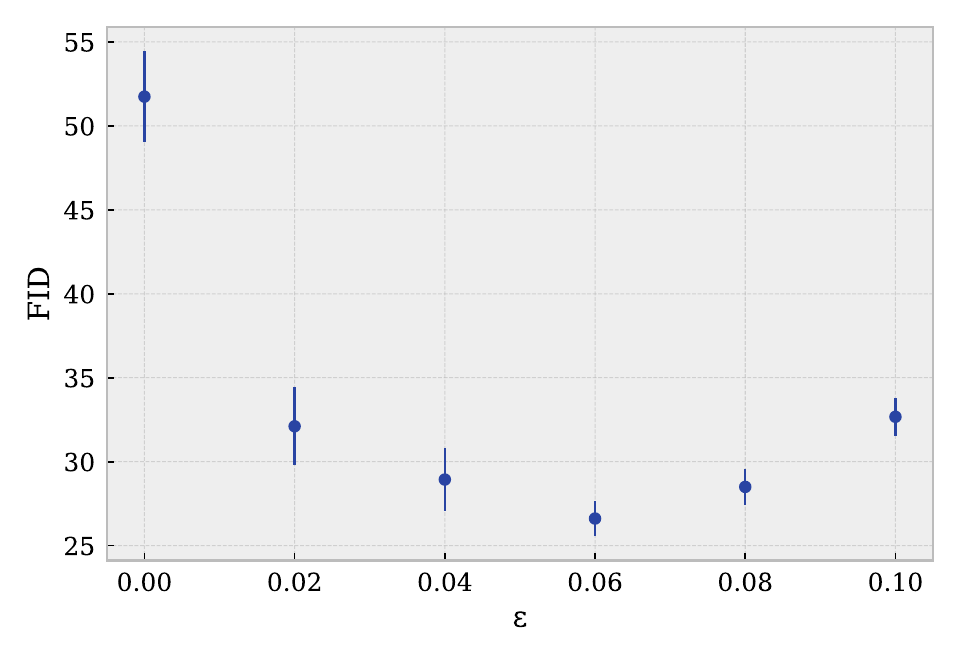}
    \caption{The Fréchet Inception Distance between the original MNIST test set (restricted to 0, 1 and 2) and a corresponding set generated with a feedforward module that receives input from the dynamical encoder. This latter set is obtained by generating the same amount of images per class as in the test set. For every value of $\epsilon$, the optimization has been repeated over ten independent realizations. }
    \label{fig:FidVsEps}
\end{figure}

To further elaborate on the role played by noise in the generative scheme, we consider the so called Fréchet Inception Distance (FID), a common metric to gauge the quality of generated images\cite{heusel2017fid}. The FID compares the distribution of generated images with the distribution of a set of real images that constitute the "ground truth" set. Mean and covariance statistics of many images as generated by the model are compared with the corresponding statistics calculated from images that belong to the reference list. Smaller FIDs point to more successful generative output. In Figure \ref{fig:FidVsEps} the computed FID for the MNIST dataset (restricted to handle 0,1,and 2) is plotted against the value of the parameter $\epsilon$, as employed in the generative Continuous-Variable Firing Rate model. As it can be clearly seen, a minimum is displayed in correspondence of an optimal value of the control parameter $\epsilon$. A controlled noise injection during training hence improves the quality of the generated images. Furthermore, the injected noise proves also beneficial in terms of computed loss (data not shown). This latter reaches its minimum for a value of $\epsilon$ close to that minimize the FID.

To further challenge the proposed scheme, we tried it against CelebFaces Attributed, a more structured dataset, which is  particularly suited for facial recognition.  Images cover large pose variations, background clutter and diverse portrayed subjects. More specifically, the dataset is made of 202599 face images of various celebrities, with 10177 unique identities. Each image reports on 40 binary attribute annotation. Images are referred to 5 landmark locations. To cope with the increased level of complexity the model is modified as follows:

\begin{itemize}
    \item Both the encoding and the decoding modules are replaced with convolutional and deconvolutional neural networks, with a feed-forward layer that links the feature maps to a flattened vector of dimension 100, where the Continuous-Variable Firing Rate model takes input from.
    \item As it is customary with reconstruction tasks, the loss function is equipped with a regularization term (usually called \textit{perceptual loss} \cite{johnson2016perceptual}) that evaluates the mean squared error between the original data and the reconstructed one. These, both the former and the latter, are processed through a pre-trained VGG19  network, a type of CNN that has been shown to be effective in capturing image content style and features \cite{simonyan2015vgg}. The mean squared distance is computed for several layers of the CNN, each corresponding to an increasing feature depth. We follow a rule of thumb of setting the coefficient in front of the regularization term such that the VGG contribution is of the same magnitude as the other losses.
\end{itemize}

\begin{figure}[!htbp]
    \centering
    \includegraphics[width=1.\linewidth]{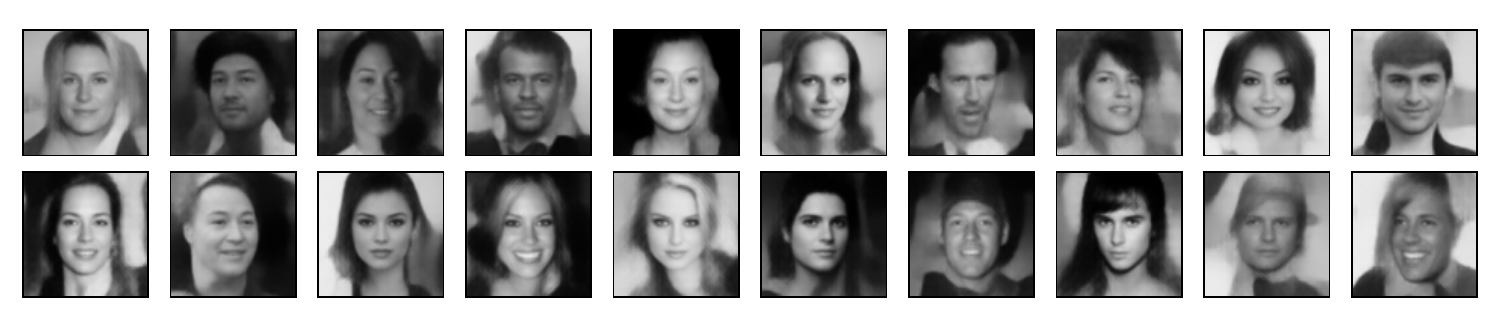}
    \caption{A gallery of generated images are displayed. The generation scheme is described in the main body of the paper. The Continuous-Variable Firing Rate model bridges the gap between the output of the encoded and the distribution in the latent space, which gets sampled by the decoder.}
    \label{fig:generated_phases}
\end{figure}

A sample of generated images is reported in Figure \ref{fig:generated_phases}. These are all faces non contained in the original dataset and that testify on the ability of the pseudo-dynamical scheme to handle the requested task. More concretely, we chose to operate with just one attractor and sampled in generation mode the distribution produced by the dynamical update rule. This latter is approximated as a multivariate Gaussian centered at the attractor location and with an empirical covariance matrix $\Sigma^{(\ell)}_T$.
The quality of the generated samples is satisfying. The fact that the background is in generally poorly resolved has an obvious cause: only few landscape (5 in total) alternate across the different images that compose the examined dataset.

\begin{figure}[!htbp]
    \centering
    \includegraphics[width=0.8\linewidth]{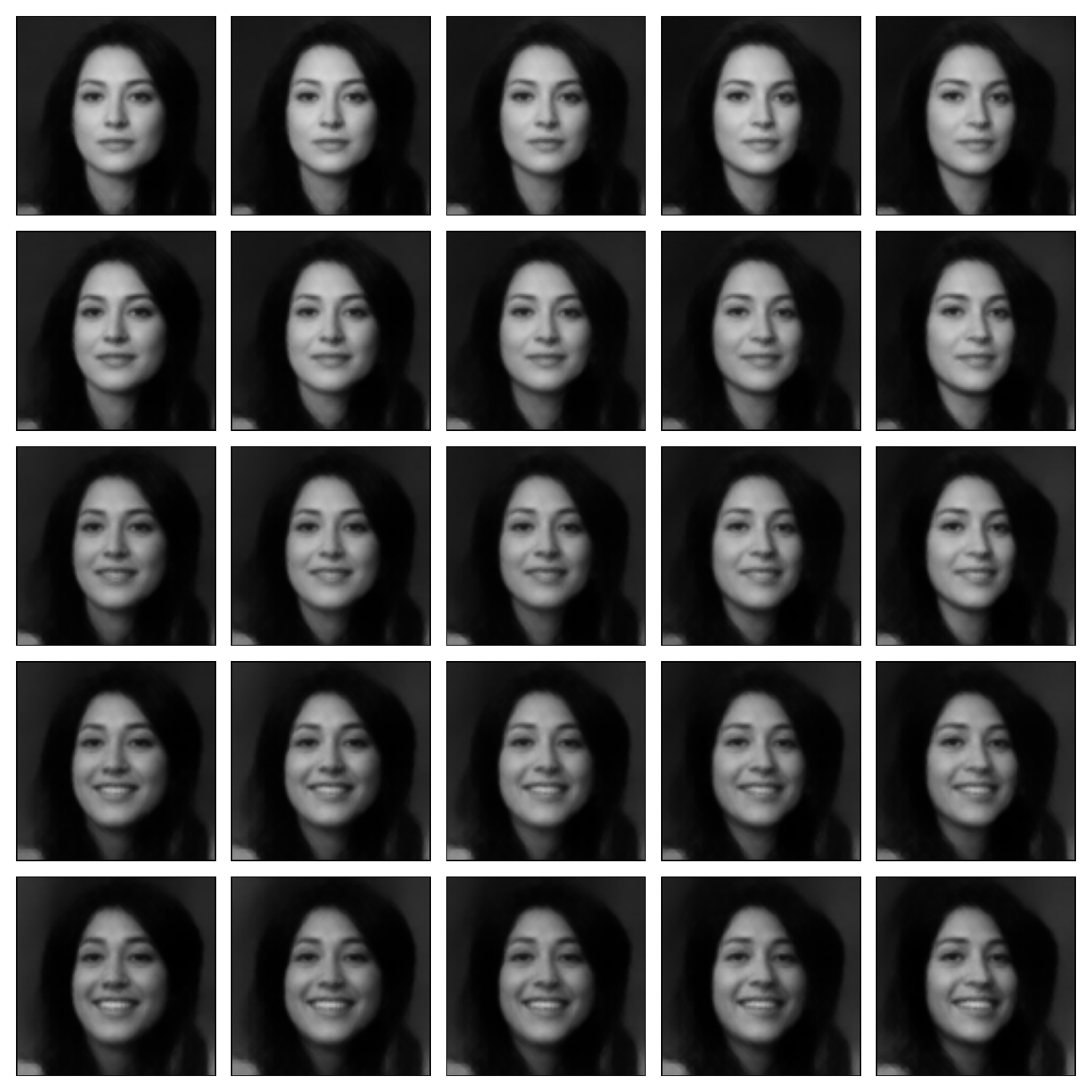}
    \caption{An example of disentanglement where we brighten a smile on an axis and rotate the face on the other axis.}
    \label{fig:disentanglement}
\end{figure}

Building on the above discussion, as carried out for the simplified MNIST framework, we expect that the dynamical module creates probability 'bubbles' in the latent space. These clouds of probability overlap when they tend toward the point attractor, sharing distinct features from different images. Such a dynamical superposition sits at the core of the generative ability. Crawling from one region to the other in the latent space, with an ad hoc modulation of the individual components, make it possible to disentangle individual features. In Figure \ref{fig:disentanglement}, starting from one input face (top left corner) we brighten the smile (when moving vertically) and induce a rotation of the images subject (when moving horizontally).

\begin{figure}
    \centering
    \includegraphics[width=1.\linewidth]{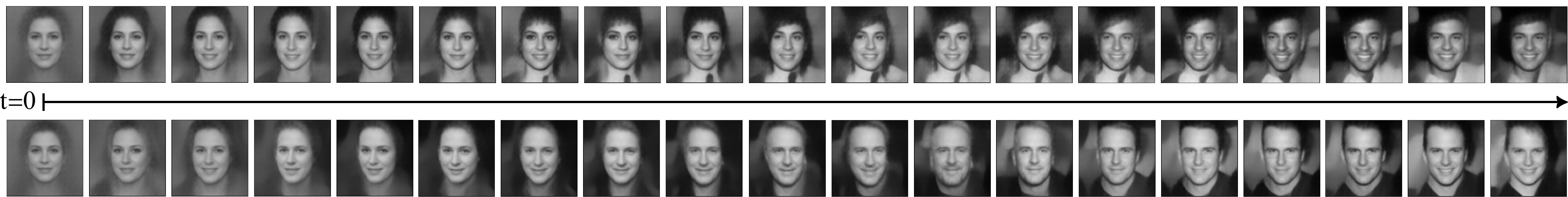}
    \caption{The left most figure (for each row) is the image that one obtains by decoding the point attractor. Starting from this initial condition, the stochastic Continuous-Variable Firing Rate model is integrated by assuming a form of the noise $G^{(\ell)}_T$ that solves the stationary Lyapunov equation for $\Sigma=\Sigma^{(\ell)}_T$. Each dynamical step of the stochastic model yield a different image, as unwrapped by the feedforward decoder. The two rows refer to two distinct realizations of the stochastic Langevin dynamics.}
    \label{Fig:langevin_sampling}
\end{figure}

An interesting follow up of the outlined generative scheme goes as follows. Imagine you are willing to generate images that populate the multivariate normal distribution $\mathcal{N}( \bar{x}^{(\ell)},\Sigma^{(\ell)}_T)$. One could in principle proceed as follows: (i) use $\Sigma^{(\ell)}_T$ in the aforementioned Lyapunov equation and estimate matrix $G^{(\ell)}_T$ that solves it at equilbrium, given the form of the Jacobian; (ii) run a Continuous-Variable Firing Rate model with a structure of the noise given by $G^{(\ell)}_T$ as estimated above. By initializing the dynamics on the deterministic attractors will force the system to visit the distribution of the latent space. Each step of the stochastic Continuous-Variable Firing Rate model could be therefore used as a possible input for the feedforward decoder to yield an original image output. The effectiveness of this generative strategy, which in extreme synthesis implements a Langevin sampling under the linear noise approximation is confirmed by inspection of Figure \ref{Fig:langevin_sampling}. Dynamical persistence makes the system swiftly pass for one image to another, distinct from the formed, through a sequence of successive adjustments.

\subsection{A dynamical model for the generative step that follows statistical sampling}

\begin{figure}
    \centering
    \includegraphics[width=1.\linewidth]{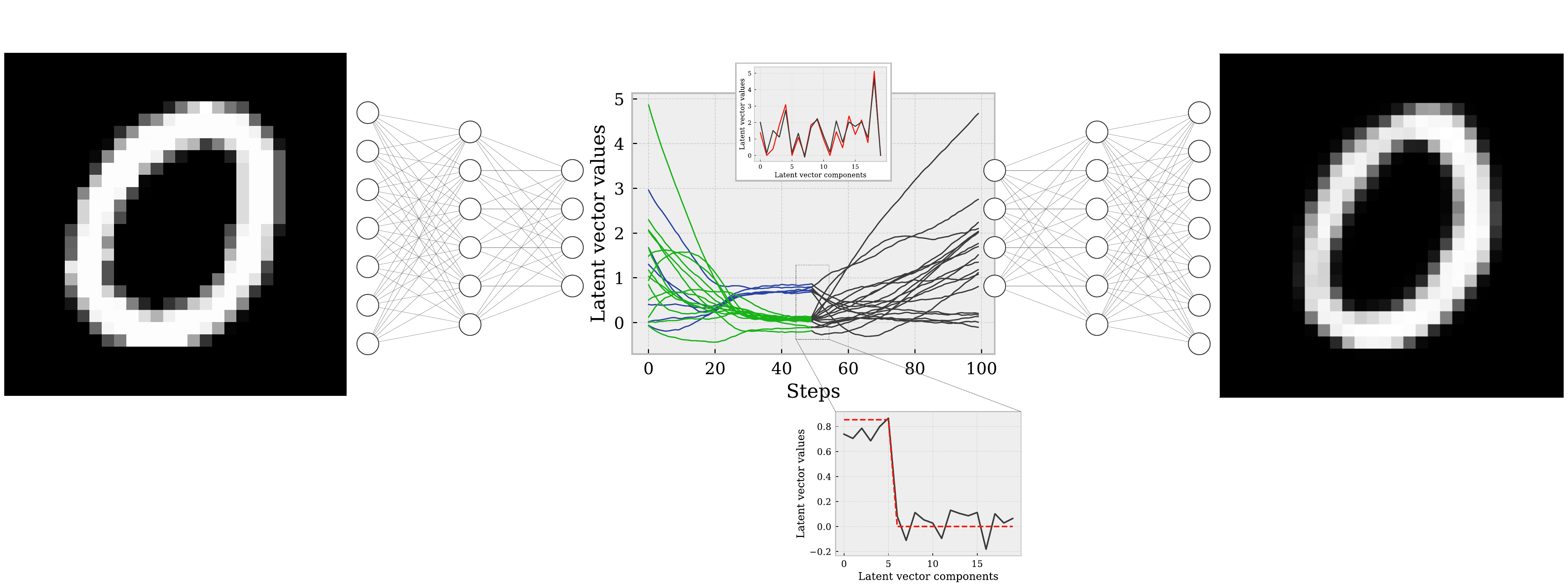}
    \caption{An image representing a number zero is processed by a (pretrained) encoder and fed as an input to the Continuous-Variable Firing Rate model. The colored trajectories (blue vs. green) point to the correct destination target, implying that the relevant attractor has been faithfully approached (see lower inset). The evolution of the first dynamical system at time $T$ is provided as an input to a second dynamical system, of the same typology. The following evolution, up to a pre-assigned time $T'$, shapes the output vector that is give as an input to the (pretrained) decoder. The produced imaged is an accurate representation of the input zero. The compressed versions ($d=20$)
  of the input and output zeros, as bridged by the cascade of the two nested dynamical system are confronted in the upper panel and they superpose quite nicely.}
  \label{dyn_gen}
\end{figure}

As anticipated above we now proceed to generalize the above recipe by replacing the decoder phase with yet another version of the stochastic Continuous-Variable Firing Rate model, characterized by a different matrix of inter-nodes coupling ($A'$), as compared to that employed in sculpting the latent probability distribution ($A$). In this latter model will posit $r_i=0$, which amounts to dropping the linear decay factor. As a proof of principle, we are back to considering MNIST restricted to digits 0, 1 and 2. For practical reasons, data are compressed in a space of dimension $d=20$ following a pretrained encoder/decoder filter. To state it clearly, the encoder and decoder modules are trained independently from the dynamical units, at variance with what implemented above. Then, the compressed data are provided as an input to the first stochastic Continuous-Variable Firing Rate model: the elements of the adjacency matrix $A$ are free parameters to be adjusted upon optimization. The output at time $T$ is fed as an entry of second Continuous-Variable Firing Rate model: the coupling matrix $A'$ define the target of the optimization. The goal is to train both $A'$ and $A$ in such a way that the convergence and reconstruction losses get minimized. Here the stochastic drive on each pixels is assumed uncorrelated ($G=G'=\mathbb{I}$). The results of the trained reconstruction algorithm are displayed in Figure \ref{dyn_gen}, for the case of an individual zero digit. Notice that both the convergence towards the probabilistic attractors, and the subsequent generation scheme are implemented via tailored stochastic model of the Continuous-Variable Firing Rate type.

\begin{figure}
    \centering
    \includegraphics[width=1.\linewidth]{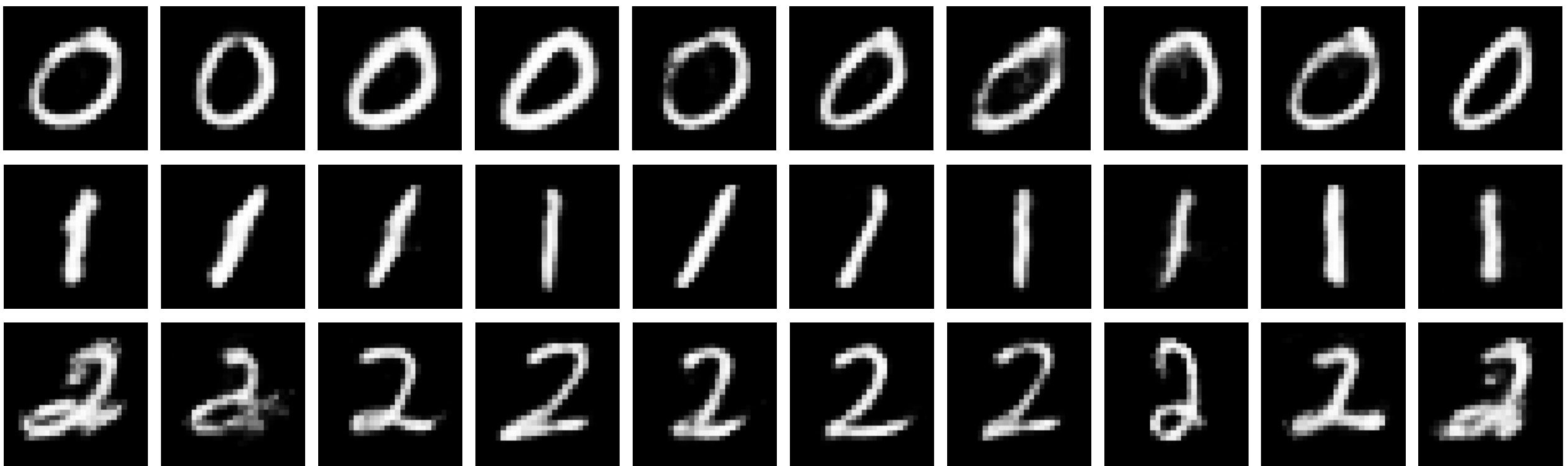}
    \caption{Example of MNIST (0, 1 and 2) generations for the fully dynamical setup.}
    \label{dyn_gens}
\end{figure}

In Figure \ref{dyn_gens} a few examples of digits generated with the full dynamical recipe are displayed. As for the mixed model that implemented a feedforward neuronal network for the decoding phase, we have here opted for a conditional generation: different numbers are indeed associated to distinct attractors. The analysis can be in principle extended, beyond this prove of principle, to more complex dataset as those considered in the preceding sub-section. This is left for future applications.

\section{Conclusions}

In this paper we have elaborated on a general approach to operate dynamical models for classification and generation purposes. To this end, we have introduced a reference dynamical model which implements (i) local reaction rules, at each node's location, and (ii) binary exchanges between nodes, paired together via existing weighted bridges. These latter are made explicit by the associated adjacency matrix, which constitutes one of the target of the optimization problem. Working with the determinist model, we have devised a procedure to plant asymptotic stationary attractors of the non linear dynamics. These latter flag for the different classes to be eventually categorized. Optimizing the dynamical system amounts to adjust the matrix of the intertwined couplings to steer the evolution towards the deputed equilibrium, depending on the specificity of the supplied initial condition, namely the item to be identified. The stochastic version of the model turns each point attractors into a probability equilibrium distribution. This latter can be analytically accessed under the linear noise approximation yielding a multivariate Gaussian solution with covariance matrix set by the corresponding Lyapunov equation. The structure of the noise that shakes the deterministic dynamics can be self-consistently learned by deploying the discretized Euler-Maruyama version of the continuous model on a deep residual network and running a backpropagation through time. The addition of noise proves beneficial in that it provides a flinch of robustness to oppose adversarial attacks. This is a non negligible effect that bears immediate reflex in the structure of the trained adjacency matrix and, as such, it is also inherited by the deterministic counterpart of the optimized stochastic model. 

Moving forward in the analysis, we also considered a variant of the model with generative capabilities. The architecture, partly inspired to Variational Autoencoders, maps each input data point into a coherent probability distribution, at time $T$, large but not infinity. Since the process is driven by a convergent stochastic model, operated under the assumption of a small noise inclusion, the dynamics is bound to asymptotically approach a multivariate Gaussian distribution. At time $T$, data are already distributed around their reference attractor, organized in adjacent 'bubbles' of probability, which provide the optimal packaging for an efficacious generation strategy. Sampling the target distribution and using the information as an input to a feedforward neural network (or even to a different dynamical model of the same typology), enables one to generate, upon training, novel data points which were not contained in the provided reservoir of examined original images. Conditional generation is also possible by combining classification and generation mode. Automatic disentanglement of isolated key features is demonstrated. This is a byproduct of the effective information storage in contiguous clusters of probability, as carved by the trained convergent dynamics.

Summing up, we have here taken one step forward in the direction of 
training a set of coupled ODEs to handle non trivial tasks, like classification (by extending on previous analysis) and generation (a thoroughly novel challenge). The ability of the trained model to cope with the assigned tasks is engraved in the architecture of the emerging network as well as in the structure of noise correlations, both target of the training process. The overall idea is to progressively replace implicit feedforward modules, with transparent dynamical units and bridge the gap between current neural network models and the field of computational neuroscience. In this work we consolidate this objective with reference to the classification task, by extending the realm of applications of previous analysis to a general class of models and elaborating further on the role of the injected noise. We also provide a first prove of principle of the possibility to assemble a fully dynamical platform for generation purposes, pretrained residual feedforward modules being solely operated for data compression in the final presented application designed to handle the restricted MNIST case study. Extending the analysis to tackle complex dataset remains a challenge that we will face in future work.

\appendix
\section{On the attractors' degeneracy}
\label{App1}

We go back to discussing the strategies for planting the attractors in the considered deterministic model. A slight improvement (that still maintains a fair separation among distinct attractors) as compared to the setting given by condition (\ref{AttrChoice}) can be straightforwardly accomplished by setting
\begin{equation}
    \bar{x}^{(\ell)}_i=\begin{cases}
        a_-\ \ \ \text{if }\ \   i=(\ell-1)L+1\\
        a_+\ \ \ \text{if }\ \   i=(\ell-1)L+2, \dots, \ell L\\
        0 \ \ \ \ \ \text{otherwise}, 
    \end{cases}
\label{AttrChoice1}   
\end{equation}
or any of the above combinations, which lowers to $2^C$ the number of identified attractors. Furthermore, one can drop the equidistance requirement and fix $2^{C'}$ stationary points with
\begin{equation}
    C' = \lceil\log_2C\rceil,
\end{equation}
following the recipe presented above.

{
Nonetheless, it should be further noted that among the spurious attractors in \eqref{spurious}, the ones with $c_\ell=a_-/a_+$ for some $\ell$ are always unstable. Intuitively, they should not undermine the effective functioning of the trained model (which leverages on a subset of stable asymptotic attractors). To prove the above claim, we set to consider the spurious attractor given by:
\begin{equation}
    \psi^{}_i=\begin{cases}
        a_-\ \ \ \text{if }\ \   i=(\ell-1)L+1, \dots, \ell L\\
        0 \ \ \ \ \ \text{otherwise}.\label{AttrRecipe}
    \end{cases}
\end{equation}
With reference to the Continuous-Variable Firing Rate model, the Jacobian evaluated on the above point is 
\begin{equation}
    J_{ij}=-\delta_{ij} +\frac{2c}{\lambda\beta a_-}\tilde{A}_{ij},
\end{equation}
where $\tilde{A}_{ij}$ is equal to $A$ for the columns $(\ell-1)L+1, \dots, \ell L$, and is otherwise null. Interestingly enough, $\psi$ is an eigenvector of $J$ with eigenvalue
\begin{equation}
    -1 +\frac{2c}{\beta a_-}.
\end{equation} 
Furthermore, at least $N-L$ eigenvectors exist with eigenvalue $-1$. While the latter favor stability, the former is positive for $c=1/8$ and $\beta=1$ (the chosen setting for the examined Continuous-Variable Firing Rate model), thus rendering every spurious attractors of the said type unstable. The same reasoning applies for any attractor built by means of $a_-$.}

\begin{figure}
    \centering
    \includegraphics[width=0.5\linewidth]{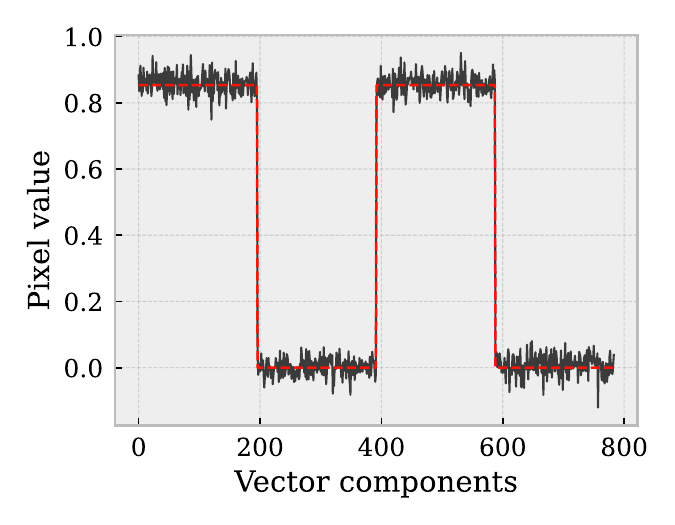}
    \caption{The red profile refers to a spurious attractor, computed as a superposition of two planted attractors. This is used to flag one of the digits' class of the MNIST database. The black curve refers to the asymptotic evolution of the generated trajectory. This latter is generated by starting from one of the images belonging to the class of pertinence, as an initial condition to the Continuous-Variable Firing Rate model. To classify 10 distinct digits class, only 4 eigenvectors should be explicitly planted.}
  \label{fig:spurious}
\end{figure}

We now proceed further by showing that spurious attractors emanating from recipe (\ref{AttrChoice1}) can be used for classification purposes. Consider for this purpose the full MNIST dataset. To identify ten distinct class of digits (numbers ranging from $0$ to $9$), only four eigenvectors are explicitly planted using \eqref{AttrRecipe}, with $L=196$. In Figure \ref{fig:spurious} 
the red line stands for the chosen spurious attractors (computed as a linear superposition of two planted attractors). The blue curve refers to the integrated state of the generated trajectory. It aligns with the crafted spurious attractor, thus correctly flagging for the class of reference. 

\begin{figure}
    \centering
    \includegraphics[width=0.5\linewidth]{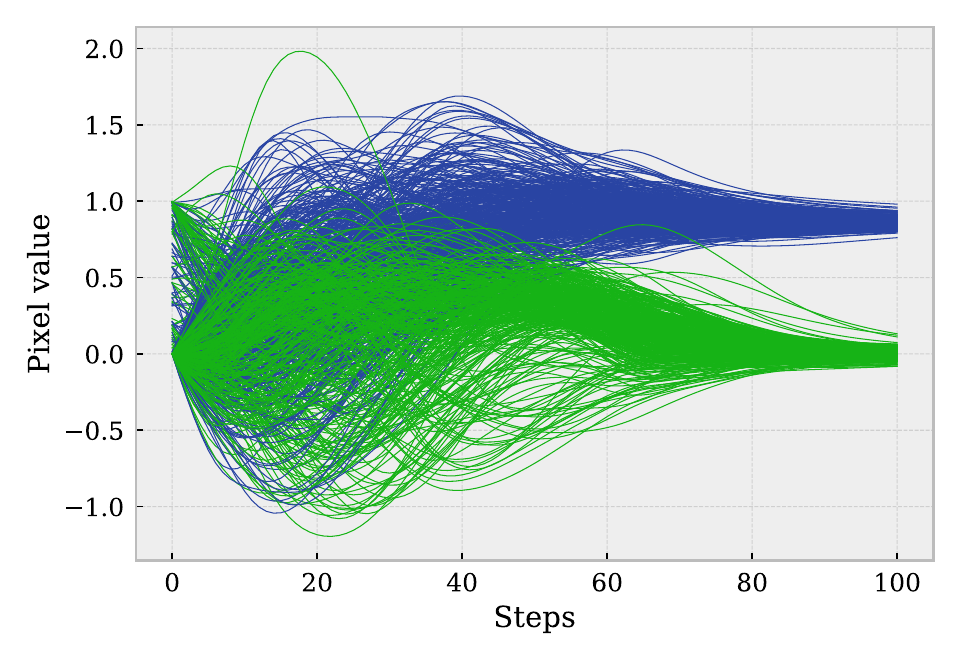}
    \caption{The evolution is plotted against time for the same example as reported in Figure \ref{spurious}. Different trajectories are plotted by using a color code which identifies the final state that the system is expected to reach for a proper classification.}
    \label{fig:spurious1}
\end{figure}

In Figure \ref{fig:spurious1} the corresponding trajectory is displayed against time. Different pixels head correctly towards the expected destination values, as identified by the corresponding color code.  

\section{On the Mahalanobis distance among trained stochastic attractors}
\label{App2}

\begin{figure}
    \centering
    \includegraphics[width=0.5\linewidth]{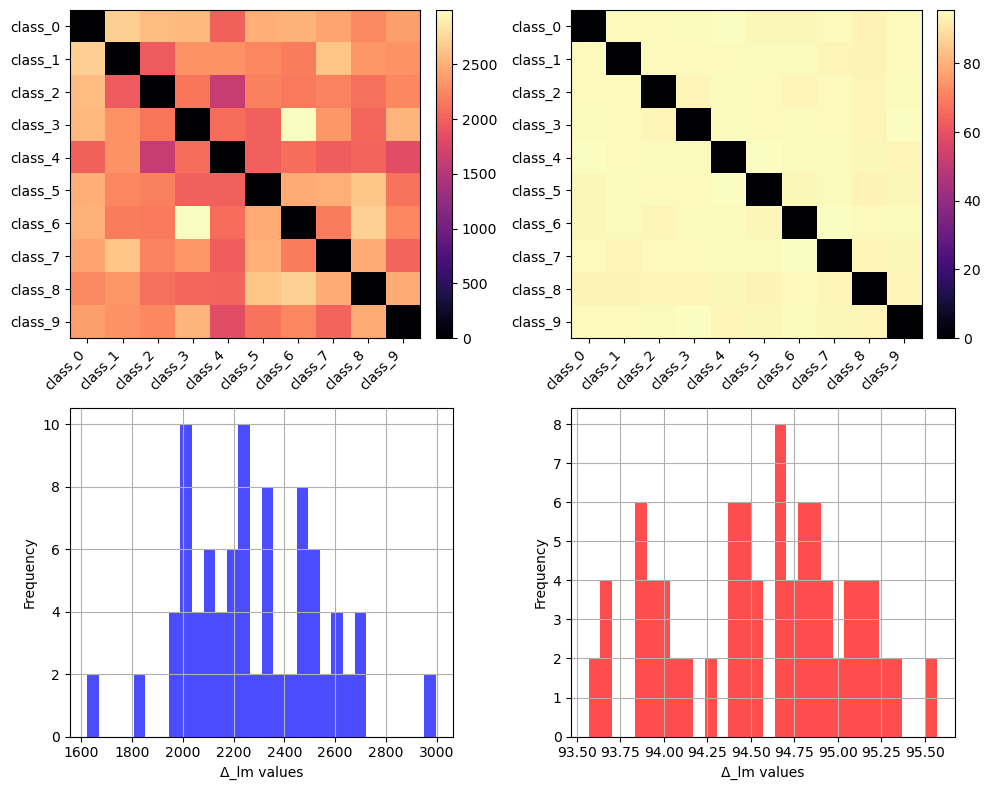}
    \caption{Comparison between model with a trainable G matrix and model with effective epsilon. The effective epsilon is calculated as: $\epsilon = \sqrt{\frac{Tr(GG^T)}{n}}$.}
    \label{fig:Mahalanobis}
\end{figure}

As anticipated in the main body of the paper, for the trained-$G$ model we observe consistently larger Mahalanobis distances between attractors. The results of the analysis as obtained for a Continuous-Variable Firing Rate model trained against the original MNIST database, are reported in Figure \ref{fig:Mahalanobis}. To draw a fair comparison, the results obtained for the setting of interest with $G$ fully trainable are confronted with those obtained by constraining $G$ to the identity matrix and strength of the noise given by $\epsilon = \sqrt{\frac{Tr(GG^T)}{n}}$. As already remarked, 
the attractors appear to be more separated when $G$ is freely modulated by the self-consistent optimization protocol as compared to what one gets by assuming uncorrelated noise of similar amplitude (as estimated ex post by calculating $\epsilon = \sqrt{\frac{Tr(GG^T)}{n}}$ ).

{\bf Acknowledgement}

The work of L.C., D. F., D. F. and R. M. is supported by \#NEXTGENERATIONEU (NGEU) and funded by the Ministry of University and Research (MUR), National Recovery and Resilience Plan (NRRP), project MNESYS (PE0000006) "A Multiscale integrated approach to the study of the nervous system in health and disease" (DR. 1553 11.10.2022).
The work by S. G. and D.F is supported by \#NEXTGENERATIONEU (NGEU) and funded by the Ministry of University and Research (MUR), National Recovery and Resilience Plan (NRRP),project MAPLE ("Bandi in Cascata") under FAIR initiative
\\

\bibliographystyle{apsrev4-1}
\bibliography{references}

\end{document}